\documentclass[aps,prl,twocolumn,superscriptaddress]{revtex4-2}

\usepackage[hidelinks]{hyperref}
\hypersetup{
	colorlinks,
	linkcolor={red},
	citecolor={blue},
	urlcolor={blue}
}
\usepackage{physics}
\usepackage{balance}
\usepackage{suffix}
\usepackage{mathtools}
\usepackage[utf8]{inputenc}
\usepackage{booktabs}
\usepackage{cases}
\usepackage[multiple]{footmisc}
\usepackage{dcolumn}
\usepackage{color,soul}
\usepackage{rotating}
\usepackage{perpage}
\usepackage{siunitx}
\usepackage{xcolor}
\usepackage{soul}
\usepackage{amsmath}
\usepackage{tikz}
\usepackage[T1]{fontenc}
\usepackage{etoolbox}
\usepackage{graphics}
\usepackage{siunitx}
\usepackage{float}	
\usepackage{collref}
\usepackage{multirow}
\usepackage{mathtools}
\usepackage{bm}
\usepackage{url}
\usepackage{pifont}
\usepackage{balance}

\usepackage{amssymb}
\usepackage{mathtools}
\usepackage{amsmath}
\usepackage{bm}
\usepackage{siunitx}
\usepackage{graphicx}
\usepackage{float}
\usepackage{multirow}
\usepackage{booktabs}
\usepackage{tikz}
\usepackage{tikz-3dplot}
\usepackage{pgfplots}
\pgfplotsset{compat=1.18}

\usepackage{tikz}
\usepackage{tikz-3dplot}
\usepackage{accents}

\usepackage{ulem}
\makeatletter
\newcommand{\doublewidetilde}[1]{{%
		\mathpalette\double@widetilde{#1}%
}}
\newcommand{\double@widetilde}[2]{%
	\sbox\z@{$\m@th#1\widetilde{#2}$}%
	\ht\z@=.9\ht\z@
	\widetilde{\box\z@}%
}
\makeatother

\usepackage{etoolbox,lipsum}

\newcommand{\blue}[1]{\textcolor{black}{#1}}

\begin{document} 
	\title{Giant perpendicular Edelstein polarization in 2D compensated magnets\\ via bichromatic Floquet driving}
	\author{Mohsen Yarmohammadi}
	\email{mohsen.yarmohammadi@georgetown.edu}
	\address{Department of Physics, Georgetown University, Washington DC 20057, USA}
	\author{Daegeun Jo}
	\affiliation{Department of Physics and Astronomy, Uppsala University, P.O. Box 516, SE-75120 Uppsala, Sweden}
	\author{Marco Berritta}
	\affiliation{Department of Physics and Astronomy, Uppsala University, P.O. Box 516, SE-75120 Uppsala, Sweden}
	\author{Libor \v{S}mejkal}
	\address{Max Planck Institute for the Physics of Complex Systems, N\"othnitzer Str.\ 38, 01187 Dresden, Germany}
	\address{Max Planck Institute for Chemical Physics of Solids, N\"othnitzer Str.\ 40, 01187 Dresden, Germany}
	\address{\mbox{Institute of Physics, Czech Academy of Sciences, Cukrovarnick\'a 10, 162 00 Praha 6, Czech Republic}}
	\author{James K. Freericks}
	\address{Department of Physics, Georgetown University, Washington DC 20057, USA}
	\author{Peter M. Oppeneer}
	\affiliation{Department of Physics and Astronomy, Uppsala University, P.O. Box 516, SE-75120 Uppsala, Sweden}
	\date{\today}
	\begin{abstract}
		While unconventional $p$-wave magnets can generate nonrelativistic Edelstein polarizations, spin-group symmetries strictly forbid these responses in unconventional magnets with higher-order harmonics, such as $d$-wave altermagnets. Here, we demonstrate that combining Rashba spin-orbit coupling with bichromatic Floquet driving activates giant perpendicular Edelstein polarizations (PEPs) across 2D altermagnets and broader classes of unconventional spin-polarized magnets—a feat monochromatic driving cannot achieve. By dynamically breaking two-fold rotational symmetry, the two-frequency drive~(including bilinear, bicircular, and circular-linear configurations) induces a stray-field-free in-plane Zeeman-like field that generates orbitally dominated PEPs (0.5--1.5 $\mu_{\rm B}$). This massive response is governed by universal selection rules tied to the system's magnetic parity and the second beam's harmonics. These emergent PEPs provide a powerful mechanism for perpendicular memory writing. 
	\end{abstract}
	
	\maketitle
	{\allowdisplaybreaks
		
		\blue{\textit{Introduction}}---Unconventional magnets (UMs), including altermagnets, host momentum-dependent spin and orbital textures protected by crystal and spin symmetries~\cite{doi:10.1126/sciadv.aaz8809,hayami2019momentum, PhysRevX.12.031042, PhysRevX.12.040501, bai2024altermagnetism, liu2025different}, enabling large band splittings~\cite{krempasky2024altermagnetic, lee2024broken, osumi2024observation} and anomalous spin--charge responses~\cite{doi:10.1126/sciadv.aaz8809,naka2019spin, mazin2021prediction, shao2021spin, Feng2022, gonzalez2021efficient, PhysRevLett.128.197202, karube2022observation} without stray fields despite zero net magnetization. Charge-to-spin conversion, such as the linear Edelstein effect \cite{edelstein1990spin}, offers a direct route to act on the spin and orbital textures. While DC currents can generate nonrelativistic Edelstein polarizations in $p$-wave magnets, rigid symmetries strictly forbid these polarizations in other $d$-, $f$-, and $g$-wave classes~\cite{hellenes2024pwavemagnets, Chakraborty2025, k9p4-tfhd}. 
		
		
		While in-plane DC electric fields routinely generate in-plane Edelstein polarizations in 2D UMs with Rashba spin-orbit coupling~(RSOC)~\cite{bychkov1984, manchon2015new,PhysRevB.93.195440,condmat10010015}, two-fold rotational symmetry around the $z$-axis ($\mathcal{C}_{2z}$) strictly forbids perpendicular Edelstein polarizations (PEPs) in equilibrium [Fig.~\ref{f1}, left column], with the notable exception of the $p$-wave systems as mentioned above. Consequently, utilizing these conventional in-plane spin responses to switch perpendicular magnetic bits necessitates external magnetic fields. This requirement poses a major hurdle for scalable magnetic random-access memory architectures~\cite{shao2021roadmap,Kang2025,Nguyen2024,clzw-1-2-022201,RevModPhys.91.035004}, which inherently rely on deterministic, field-free switching aligned with the perpendicular magnetic anisotropy.
		
		PEPs overcome this limitation because these out-of-plane spin or orbital accumulations couple directly to the magnetic order, enabling field-free current-induced torques~\cite{shao2021roadmap,Kang2025,wang2023field,kao2022deterministic}. This is particularly attractive for compensated magnets, where zero net magnetization suppresses stray fields while momentum-dependent spin splitting sustains sizable nonequilibrium polarizations. Together with recent advances in orbital Edelstein effects and orbital torques~\cite{Salemi2019,fukunaga2023orbital,yang2024orbital,ding2024orbital}, controlled spin and orbital PEPs provide a route toward low-dissipation spintronic and orbitronic functionalities.\begin{figure}[t]
			\centering
			\includegraphics[width=0.95\linewidth]{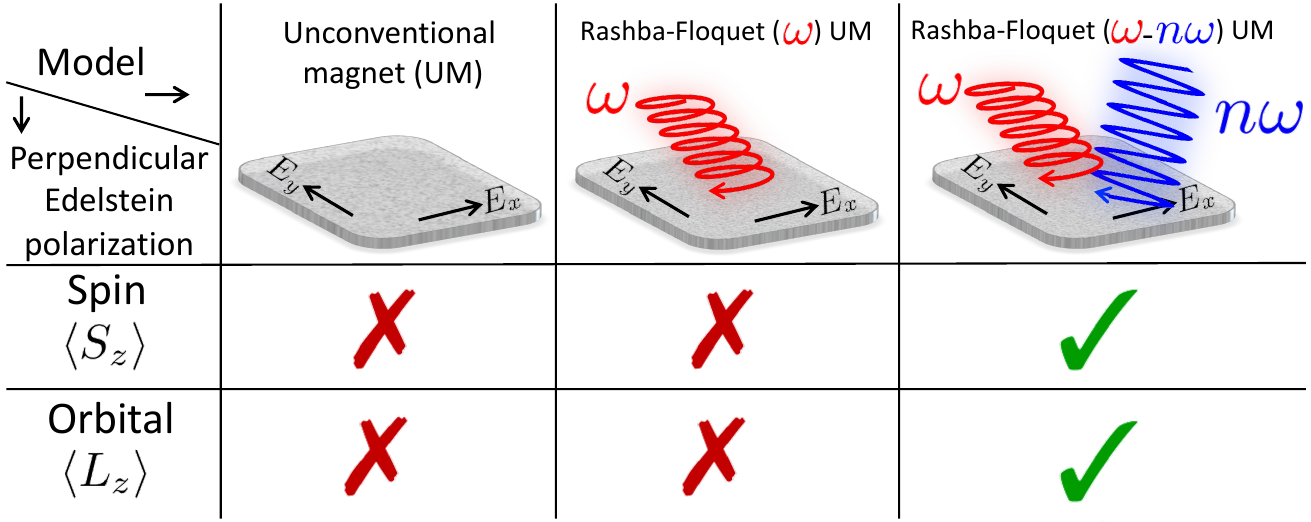}
			\caption{Schematic of out-of-plane spin ($\langle S_z\rangle$) and orbital ($\langle L_z\rangle$) Edelstein polarizations in a 2D Rashba UM induced by in-plane DC electric fields $(E_x,E_y)$ and different drives. In equilibrium and under monochromatic driving, out-of-plane responses are forbidden by $\pi$-rotation $\mathcal{C}_{2z}$ symmetry. Bichromatic driving instead generates finite $\langle S_z\rangle$ and $\langle L_z\rangle$ by dynamically breaking $\mathcal{C}_{2z}$, with a strong dependence on magnetic parity and driving (bi-linear, bi-circular, and mixed beams) protocol.}
			\label{f1}
		\end{figure}
		
		Static approaches to PEPs via strain, gating, or magnetic fields~\cite{fert2024electrical,Dieny2020,Ji2024,Yu31122022,Zunger2021,Matsukura2015,Barla2021} suffer from limited tunability, CMOS integration hurdles, and lattice instabilities. Furthermore, as noted above, purely electrical DC routes remain completely prohibited by rigid point-group symmetries in UMs with out-of-plane spin orientations~\cite{golub2025spinorientationelectriccurrent, din2025unconventionalrelativisticspinpolarization, campos2026persistentaltermagnetism}. While contactless, ultrafast laser driving~\cite{freimuth2021laser,kimel2019writing} bypasses these static limitations, conventional single-color driving in 2D Rashba UMs generates a $\mathcal{C}_{2z}$-preserving out-of-plane Zeeman-like field $\mathcal{B}_z$~\cite{xt23-9pnv}, which does not allow induced
		PEPs [Fig.~\ref{f1}, middle column].
		
		In this Letter, we show that generalized two-frequency Floquet driving ($\omega$--$n\omega$)~\cite{yarmohammadi2026floquetinducedanisotropicmagnetoresistanceanomalous,yarmohammadi2026efficienttwocolorfloquetcontrol} overcomes these static constraints by providing a dynamic, all-optical pathway to actively break the limiting $\mathcal{C}_{2z}$ symmetry. This fundamentally unlocks the ability to generate PEPs in the otherwise prohibited $p$-, $d$-, $f$-, and $g$-wave classes~(see below for the specific $p$-wave model employed in this work). While true bichromatic driving ($n \ge 2$) is required for higher-order classes, we include the $n=1$ case—representing monochromatic elliptical driving—to cover lower-order ones for the sake of completeness. We show that nonlinear interference between the two optical frequencies generates a tunable, in-plane Zeeman-like field $\boldsymbol{\mathcal{B}}_\parallel=(\mathcal{B}_x,\mathcal{B}_y)$ mediated by the interplay of RSOC and the intrinsic magnetic parity. Because this field is generally misaligned with the principal anisotropy axes, it explicitly breaks $\mathcal{C}_{2z}$ symmetry, allowing for finite, on-demand PEPs [Fig.~\ref{f1}, right column]. The exact selection rules governing this effect depend heavily on both the magnetic parity and the driving protocol.
		
		\blue{\textit{Model}}---We consider a 2D Rashba UM near the $\Gamma$ point, described by the low-energy continuum Hamiltonian~\cite{PhysRevX.12.031042,PhysRevX.12.040501,5ys6-pq68,zt4l-y18j,PhysRevB.111.125420}\begin{align}\label{eq_1}
			H_\ell(\mathbf{k}) = \varepsilon k^2 \sigma_0 + \lambda_{\rm R}(k_x\sigma_y - k_y\sigma_x) + \varepsilon M_{\mathbf{k}}^\ell\sigma_z\,,
		\end{align}where $\varepsilon=\hbar^2/2m_{\rm e}$, $k^2=k_x^2+k_y^2$,  and the Pauli matrices $\boldsymbol{\sigma}=(\sigma_x,\sigma_y,\sigma_z)$ act in spin space. The term $\varepsilon M_{\mathbf{k}}^\ell\sigma_z$ dictates the momentum-dependent out-of-plane spin splitting generated by the out-of-plane magnetic order. Though defined in the $x$–$y$ plane, the model easily adapts to $x$–$z$ and $z$–$y$ UMs by appropriately rotating the in-plane DC probe fields.
		
		The momentum-dependent order parameters for representative odd-parity ($p, f$) and even-parity ($d, g$) states are given by{\small\begin{subequations}
				\begin{align}
					&M^p_{\mathbf{k}}
					=
					M^p_1 k_x + M^p_2 k_y, \quad \mapsto \quad \propto k^1\\
					&M^d_{\mathbf{k}}
					=
					M^d_1\left(k_x^2-k_y^2\right)
					+
					M^d_2 k_x k_y , \quad \mapsto \quad \propto k^2\\
					&M^f_{\mathbf{k}}
					=
					M_1^f k_x\left(k_x^2-3k_y^2\right)
					+
					M_2^f k_y\left(k_y^2-3k_x^2\right), \quad \mapsto \quad \propto k^3\\
					&M^g_{\mathbf{k}}
					=
					M_1^g\left(k_x^4-6k_x^2k_y^2+k_y^4\right)
					+
					4M_2^g k_xk_y\left(k_x^2-k_y^2\right), \mapsto \, \propto k^4
				\end{align}
		\end{subequations}}with the understanding that this approach can be readily extended to higher-order multipoles like $h$- and $i$-wave states. We parameterize the spin splitting as $M_1^\ell = M^\ell \cos(m\Theta_\ell)$ and $M_2^\ell = M^\ell \sin(m\Theta_\ell)$, with $m \in \{1,2,3,4\}$ dictating the $p$-, $d$-, $f$-, and $g$-wave cases. The angle $\Theta_\ell$ controls the modulation orientation (e.g., $\Theta_d=0$ and $\pi/4$ yield the $d_{x^2-y^2}$ and $d_{xy}$ wave patterns, respectively). Ultimately, the Rashba coupling generates a helical in-plane spin texture, while the magnetic term drives a staggered momentum-dependent spin splitting that alternates sign across the Brillouin zone to ensure a strictly zero net magnetization.
		
		Coupling $M_k^p$ to $\sigma_z$ in Eq.~\eqref{eq_1} models a $p$-wave magnet in the strong spin-degenerate hopping limit~\cite{khodas2026nonrelativisticisingsuperconductivitypwavemagnets,mitscherling2026microscopicoriginpwavemagnetism}. Unlike generic variants~\cite{hellenes2024pwavemagnets, Chakraborty2025}, the current model's equilibrium out-of-plane Edelstein response strictly vanishes, rendering it unsuitable for static perpendicular charge-to-spin conversion. Crucially, we demonstrate that combining RSOC with multi-color Floquet driving actively circumvents this restriction, unlocking a giant PEP in these otherwise forbidden systems~(see results).
		
		To achieve dynamical control, the system is driven by two polarized laser fields at frequencies $\omega$ and $n\omega$, introduced via the Peierls substitution $\mathbf{k}\rightarrow \mathbf{k}+e\mathbf{A}(t)/\hbar$~\cite{doi:10.7566/JPSJ.94.111007,annurev-conmatphys-031218-013423,PhysRevResearch.4.033213,Rudner2020,Giovannini_2020,Mrudul:21}. The generalized two-color field is $\mathbf{A}(t) = (\mathcal{A}_0[\cos(\omega t)
		+ \mathcal{S}\cos(n\omega t+\phi)\cos\psi],\mathcal{A}_0[\eta_1\sin(\omega t)
		+ \eta_2\mathcal{S}\cos{\small(}n\omega t+\phi-\tilde{\phi}{\small)}\cos{\small(}\psi-\tilde{\psi}{\small)}])$. Here, $\mathcal{A}$ defines the fundamental amplitude, $\omega$ ($n\omega$) the base (harmonic) frequency, $\mathcal{S}$ the harmonic intensity ratio, and $\phi,\tilde{\phi}$ the inter-beam phase delays. Light helicity and polarization orientation are controlled by $\{\eta_1,\eta_2\}$ and $\{\psi,\tilde{\psi}\}$~(incident angles), respectively. This generic form compactly encapsulates three distinct driving regimes: (i) bilinear ($\eta_1=0$, $\eta_2=1$, $\tilde{\phi}=0$, $\tilde{\psi}=\pi/2$, $\psi\neq0$), (ii) bicircular ($\{\eta_1,\eta_2\}=\pm1$, $\tilde{\phi}=\pi/2$, $\psi=\tilde{\psi}=0$), and (iii) hybrid circular-linear ($\eta_1=\pm1$, $\eta_2=1$, $\tilde{\phi}=0$, $\tilde{\psi}=\pi/2$, $\psi\neq0$).
		
		In the off-resonant high-frequency regime ($\hbar\omega \gtrsim 1$ eV, well above the bandwidth), we use the van Vleck expansion up to $1/\omega$ to obtain a time-independent effective Hamiltonian describing the prethermal Floquet state~\cite{Bukov04032015,RevModPhys.89.011004,PhysRevB.79.081406,annurev-conmatphys-031218-013423,Rudner2020,PhysRevB.82.235114,PhysRevX.3.031005,GRIFONI1998229,PLATERO20041,SciPostPhys.20.2.059,lkf9-jgv6}. For an arbitrary $\ell$-wave Rashba UM with $k^m$ spin splitting, the effective Hamiltonian reads $H_{\ell}^{\rm eff}(\mathbf{k}) = h^{\ell}_0(\mathbf{k}) \sigma_0 + \mathbf{h}^{\ell}(\mathbf{k})\cdot \boldsymbol{\sigma}$ (see Secs.~S1--S3 of the Supplemental Materials~(SM)~\cite{SM}), with\begin{subequations}\label{eq_4}
			\begin{align}
				h^{\ell}_0(\mathbf{k}) &= \varepsilon k^2 + \mu^{\ell}_\omega,\\
				h^{\ell}_x(\mathbf{k}) &= -\lambda_{\rm R}k_y + {\textstyle \sum_{r,s \in \mathbb{Z}^{+}}} \mathcal{H}^{\ell,(r,s)}_x k_x^r k_y^s + \mathcal{B}^{\ell}_x,\\
				h^{\ell}_y(\mathbf{k}) &= ~\lambda_{\rm R}k_x + {\textstyle \sum_{r,s \in \mathbb{Z}^{+}}} \mathcal{H}^{\ell,(r,s)}_y k_x^r k_y^s + \mathcal{B}^{\ell}_y,\\
				h^{\ell}_z(\mathbf{k}) &= ~M_{\mathbf{k}}^{\ell} + {\textstyle \sum_{r,s \in \mathbb{Z}^{+}}} \mathcal{H}^{\ell,(r,s)}_z k_x^r k_y^s + \mathcal{B}^{\ell}_z.
			\end{align}
		\end{subequations}The terms $\sum_{r,s \in \mathbb{Z}^{+}} \mathcal{H}^{\ell,(r,s)}_\alpha k^r_x k^s_y$ ($\alpha=x,y,z$) encode light-induced higher-order parity channels, depending sensitively on both the laser parameters and the intrinsic parity $\ell$ of the system~\cite{SM}. The term $\mu_\omega$ is a scalar AC Stark shift, while $\boldsymbol{\mathcal{B}}=(\mathcal{B}_x,\mathcal{B}_y,\mathcal{B}_z)$ denotes the effective light-induced Zeeman field. The in-plane components $\boldsymbol{\mathcal{B}}_\parallel=(\mathcal{B}_x,\mathcal{B}_y)$ arise from interference between single- and two-photon processes of the $\omega$ and $n\omega$ drives, strictly vanishing in the single-color limit~\cite{xt23-9pnv}. Their magnitude and orientation are tunable via the phase $\phi$, polarization angle $\psi$, amplitude ratio $\mathcal{S}$, and chiralities $\eta_{1,2}$. Originating from the interplay between RSOC and magnetic order, these fields vanish for $M^\ell=0$ but play a central role in enabling the PEPs discussed below. Because the planar components are driven primarily by the bichromatic interference, the rotational symmetry $\mathcal{C}_{2z}$ dictates their contribution to the effective Hamiltonian. Under the operation $\mathcal{C}_{2z}=-i\sigma_z$, the momentum $(k_x,k_y)\!\to\!(-k_x,-k_y)$ and spin $(\sigma_x,\sigma_y)\!\to\!(-\sigma_x,-\sigma_y)$ transformations leave $\sigma_0$ and $\sigma_z$ invariant. Consequently, equilibrium and single-color Rashba-Floquet ($\omega$) UMs preserve $\mathcal{C}_{2z}$ symmetry, whereas the bichromatic $\omega$--$n\omega$ drive explicitly breaks it via the transverse spin components $(h_x(\mathbf{k}),h_y(\mathbf{k}))$.
		
		We analyze bi-linear, bi-circular, and mixed circular--linear driving for all 2D Rashba $\ell$-wave UMs and find that $\mathcal{C}_{2z}$ breaking depends sensitively on both the driving protocol and magnetic symmetry. Since $\mathcal{C}_{2z}$ violation enables DC current-induced finite PEPs, the resulting classification is summarized in Tab.~\ref{tab1}. For even-parity magnets with $k^m$ ($m=2,4,\dots$), $\omega$--$n\omega$ driving allows for PEPs when $n=m$ or $m-2$, independent of laser polarization. For odd-parity magnets ($m=1,3,\dots$), this condition holds only for bi-linear polarization, whereas bi-circular and mixed circular--linear driving generate PEPs for all $n\ge1$. Below, we focus on bi-circular driving and a $d$-wave altermagnet~(AM) as a representative case, as other configurations lead to distinct but qualitatively similar trends. Details for bi-linear and mixed cases in $d$-wave AMs are provided in~{\hypersetup{linkcolor=black}\hyperlink{mylinkB}{End Matter}}.\begin{table}[t]
			\centering
			\includegraphics[width=0.9\linewidth]{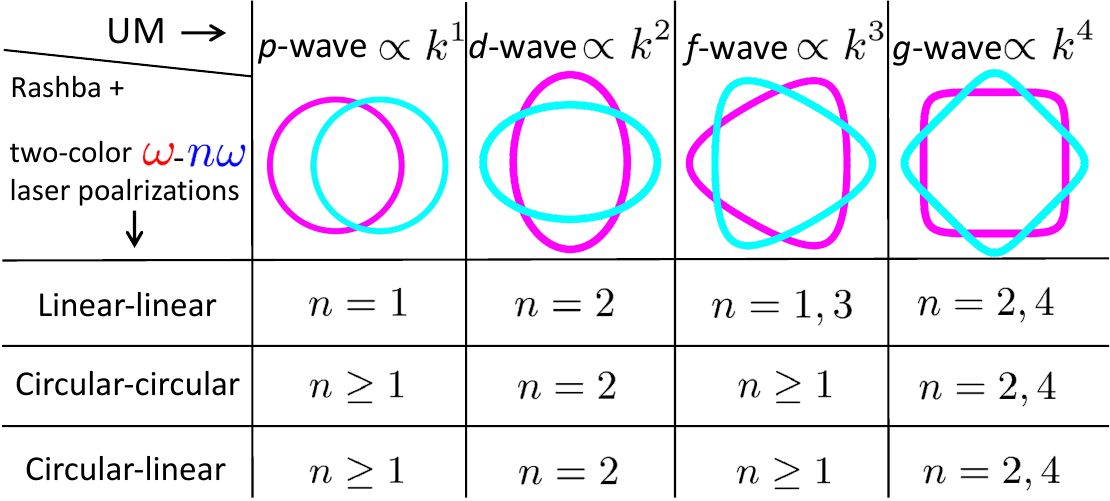}
			\caption{Required harmonic order $n$ of the secondary drive in a bichromatic (bi-linear, bi-circular, and mixed beams) $\omega$--$n\omega$ field to enable
				both spin and orbital PEP in a 2D Rashba UM. The top row illustrates the pristine characteristic momentum-space spin-splitting geometries of the resulting $p$-, $d$-, $f$-, and $g$-wave UM states, where magenta and cyan contours represent opposite spin species.}
			\label{tab1}
		\end{table}
		
		The light-driven effective Hamiltonian of a 2D Rashba $d$-wave AM with a $k^2$ spin-splitting order and two-color $\omega$--$2\omega$ scheme~[see Tab.~\ref{tab1} and Sec.~S4 of the SM~\cite{SM} for details] takes the form $H_{d}^{\rm eff}(\mathbf{k}) = h^{d}_0(\mathbf{k}) \sigma_0 + \mathbf{h}^{d}(\mathbf{k})\cdot \boldsymbol{\sigma}$ with~(see~{\hypersetup{linkcolor=black}\hyperlink{mylinkA}{End Matter}} for the derivation){\small\begin{subequations}\label{eq_4}
				\begin{align}
					h^{d}_0(\mathbf{k}) &= ~\varepsilon k^2 + \mu^{d}_\omega,\\
					h^{d}_x(\mathbf{k}) &= -\big[(\lambda_{\rm R}-\lambda_{{\rm R}\omega})k_y + \lambda_{{\rm D}\omega} k_x\big] + \mathcal{B}^{d}_x,\label{eq_4b}\\
					h^{d}_y(\mathbf{k}) &= \big[(\lambda_{\rm R}+\lambda_{{\rm R}\omega})k_x + \lambda_{{\rm D}\omega} k_y\big] + \mathcal{B}^{d}_y,\label{eq_4c}\\
					h^{d}_z(\mathbf{k}) &= ~M_{\mathbf{k}}^{d} + \mathcal{B}^{d}_z.
				\end{align}
		\end{subequations}}The light-induced parameters are $\mu^{d}_\omega = e^2 \varepsilon \mathcal{A}_0^2\left(1+\mathcal{S}^2\right)/\hbar^2$, $\lambda_{{\rm R}\omega} = e^2 \lambda_{\rm R} M_1^{d} \mathcal{A}_0^2 (2\eta_1 + \eta_2 \mathcal{S}^2)/2\hbar \omega m_{\rm e}$, $\lambda_{{\rm D}\omega} = e^2 \lambda_{\rm R} M_2^{d} \mathcal{A}_0^2 (2\eta_1 + \eta_2 \mathcal{S}^2)/2\hbar \omega m_{\rm e}$, $\mathcal{B}^{d}_x = e^3 \lambda_{\rm R} \mathcal{A}_0^3 \mathcal{S}[(1-2\eta_1\eta_2) M_1^{d}\sin\phi + (\eta_1 -2\eta_2) M_2^{d}\cos\phi]/4\hbar^2 \omega m_{\rm e}$, $\mathcal{B}^{d}_y = e^3 \lambda_{\rm R} \mathcal{A}_0^3 \mathcal{S} [(\eta_1 \eta_2-2) M_2^{d}\sin\phi + (2\eta_1-\eta_2) M_1^{d}\cos\phi]/4\hbar^2 \omega m_{\rm e}$, and $\mathcal{B}^{d}_z = -e^2 \lambda_{\rm R}^2 \mathcal{A}_0^2 (2\eta_1 + \eta_2 \mathcal{S}^2)/2\hbar^3 \omega$, where $M_1^{d} = M^{d} \cos(2\Theta_M^{d})$ and $M_2^{d} = M^{d} \sin(2\Theta_M^{d})$. In $d$-wave AMs, the Floquet commutator structure yields renormalized Rashba--Dresselhaus couplings $\lambda_{{\rm R}\omega}$ and $\lambda_{{\rm D}\omega}$. We stress that the in-plane fields $\boldsymbol{\mathcal{B}}^{d}_\parallel=(\mathcal{B}^{d}_x,\mathcal{B}^{d}_y)$ break $\mathcal{C}_{2z}$ symmetry, enabling finite PEPs while simultaneously tuning their sign and magnitude via $\omega$, $\mathcal{A}_0$, $\mathcal{S}$, $\phi$, $M^d$, $\eta_{1,2}$, and $\lambda_{\rm R}$.
		
		\blue{\textit{Edelstein polarizations}}---In inversion-asymmetric systems, an applied DC electric field $\mathbf{E}$ induces nonequilibrium spin and orbital polarizations, known as the Rashba–Edelstein effect~\cite{edelstein1990spin, Salemi2019,Johansson_2024,PhysRevMaterials.5.074407}. Within linear response, $\langle \hat{O}_\alpha \rangle = \sum_{\beta}\chi_{\alpha\beta} E_\beta$, where $\chi_{\alpha\beta}$ is the Edelstein susceptibility and $\hat{O}$ denotes either spin $\hat{\mathbf{S}}$ or orbital $\hat{\mathbf{L}}$. In the relaxation-time approximation as a baseline conceptual illustration, the Kubo formula reads~\cite{Salemi2019}{\small\begin{align}
				\chi_{\alpha \beta} = {} & \frac{ie}{m_{\rm e}} \int \frac{d^2k}{(2\pi)^2} \Bigg[ \sum_\nu \frac{\partial f_{\nu\mathbf{k}}}{\partial \epsilon_{\nu\mathbf{k}}}  \frac{O^\alpha_{\nu\nu,\mathbf{k}} p^\beta_{\nu\nu,\mathbf{k}}}{i\tau_{\rm intra}^{-1}} \notag \\ {} &- \sum_{\nu\neq \zeta} \frac{f_{\nu\mathbf{k}} - f_{\zeta\mathbf{k}}}{\epsilon_{\nu\mathbf{k}} - \epsilon_{\zeta\mathbf{k}}} \frac{O^\alpha_{\zeta\nu,\mathbf{k}} p^\beta_{\nu \zeta,\mathbf{k}}}{\epsilon_{\zeta\mathbf{k}} - \epsilon_{\nu\mathbf{k}} + i \tau_{\rm inter}^{-1}} \Bigg],
		\end{align}}where $O^\alpha_{\zeta\nu,\mathbf{k}} = \langle u_{\zeta\mathbf{k}} | \hat{O} | u_{\nu\mathbf{k}} \rangle$ and $p^\beta_{\nu\zeta,\mathbf{k}} = m_{\rm e}\langle u_{\nu\mathbf{k}} | \hat{v}_\beta | u_{\zeta\mathbf{k}} \rangle$, with $\hat{v}_\beta= \hbar^{-1}\partial H_{d}^{\rm eff}(\mathbf{k})/\partial k_\beta$. Here, $\epsilon_{\nu\mathbf{k}}$ and $u_{\nu\mathbf{k}}$ are the eigenvalues and eigenstates of the Floquet Hamiltonian $H_{d}^{\rm eff}(\mathbf{k})$ for band $\nu$, while $\tau_{\rm intra}$ and $\tau_{\rm inter}$ denote the intraband and interband lifetimes. $f_{\nu\mathbf{k}}$ is the Fermi–Dirac distribution at chemical potential $\mu$ and temperature $k_{\rm B}T$. At low temperatures, the intraband contribution dominates. In the $T\to0$ limit, the derivative of the Fermi function becomes a delta function, yielding $\chi_{\alpha \beta}^{\rm intra} = -\frac{e \tau_{\rm intra}}{m_{\rm e}} \int \frac{d^2k}{(2\pi)^2} \sum_\nu \delta(\epsilon_{\nu\mathbf{k}} - \mu)\, O^\alpha_{\nu\nu,\mathbf{k}}\, v^\beta_{\nu\nu,\mathbf{k}}.$ 
		
		For the orbital channel, the Bloch-state orbital magnetic moment follows the Berry-phase formulation~\cite{xiao2005berry}, directly related to the geometry of the Floquet bands: $ L_{\zeta\nu,\mathbf{k}} = i \langle \nabla_{\mathbf{k}} u_{\zeta\mathbf{k}} | \times [H_{d}^{\rm eff}(\mathbf{k}) - \epsilon_{\nu\mathbf{k}}] | \nabla_{\mathbf{k}} u_{\nu\mathbf{k}} \rangle.$ Importantly, this Bloch-state orbital polarization arises strictly from the self-rotation of the electron wavepacket in momentum space. It is fundamentally distinct from the atomic orbital Edelstein effect and does not require explicit atomic orbital character to manifest.
		
		The $\mathcal{C}_{2z}$ rotational symmetry strictly enforces the cancellation of any net PEP. Since this work focuses on PEPs, we restrict attention to the components $\chi_{zx}$ and $\chi_{zy}$. We note that the in-plane Edelstein susceptibilities are already finite in both Rashba AMs and Rashba--Floquet~($\omega$) AMs~\cite{xt23-9pnv}, and are instead modulated by two-color laser driving.
		
		\blue{\textit{Results and discussion}}---All responses are computed using the light-dressed effective Hamiltonian $H_{d}^{\rm eff}(\mathbf{k})$, which includes two-color driving. We set $k_{\rm B}T \approx 1~\mathrm{meV}$ and normalize the Edelstein susceptibility by $\chi_0 = e/4\pi^2$ (with $e=\hbar=m_{\rm e}=1$). Typical low-temperature scattering times in high-quality 2D materials are on the order of tens of fs; accordingly, we use $\tau_{\mathrm{intra}}^{-1}=\tau_{\mathrm{inter}}^{-1}=0.1~\mathrm{eV}$, corresponding to $ \approx 6.6~\mathrm{fs}$. The chemical potential is set near the bottom of the lower band to maximize the density of states and thus the polarization. We fix $\Theta_M^d=0$, $M_d=1$, and a weak RSOC $\lambda_{\rm R}=0.3~\mathrm{eV}$\AA, while results for other $\Theta_M^d$ and $\lambda_{\rm R}$ are provided in~{\hypersetup{linkcolor=black}\hyperlink{mylinkC}{End Matter}}.
		
		Figure~\ref{f2} shows the out-of-plane Edelstein susceptibilities versus the relative amplitude $\mathcal{S}$ of the second harmonic, separating intraband [Fig.~\ref{f2}(a)] and interband [Fig.~\ref{f2}(b)] contributions into spin and orbital parts. For single-color circular driving ($\mathcal{S}=0$), the effective Hamiltonian acquires only a uniform $\mathcal{B}_z$ shift and renormalized Rashba--Dresselhaus terms, however, $\mathcal{C}_{2z}$ is still preserved, since $\mathcal{B}_z^d$ is even under $\mathcal{C}_{2z}$, and PEPs are therefore forbidden. In contrast, bichromatic ($\omega$--$2\omega$) driving ($\mathcal{S}\neq0$) generates nonlinear interference that induces a tunable in-plane field $\boldsymbol{\mathcal{B}}_\parallel^d=(\mathcal{B}_x^d,\mathcal{B}_y^d)$, present only in the presence of altermagnetic splitting. This field, arising from Rashba–altermagnetic coupling and circular driving, breaks $\mathcal{C}_{2z}$ and enables finite perpendicular spin and orbital responses, $\langle S_z\rangle=\chi_{{\rm Spin},zx}E_x+\chi_{{\rm Spin},zy}E_y$ and $\langle L_z\rangle=\chi_{{\rm Orbital},zx}E_x+\chi_{{\rm Orbital},zy}E_y$, which are absent in both equilibrium and single-color regimes. Unlocking these polarizations relies on fundamental symmetry breaking rather than delicate parametric resonances, making their generation inherently robust against variations in the optical drive.\begin{figure}[t]
			\centering
			\includegraphics[width=0.9\linewidth]{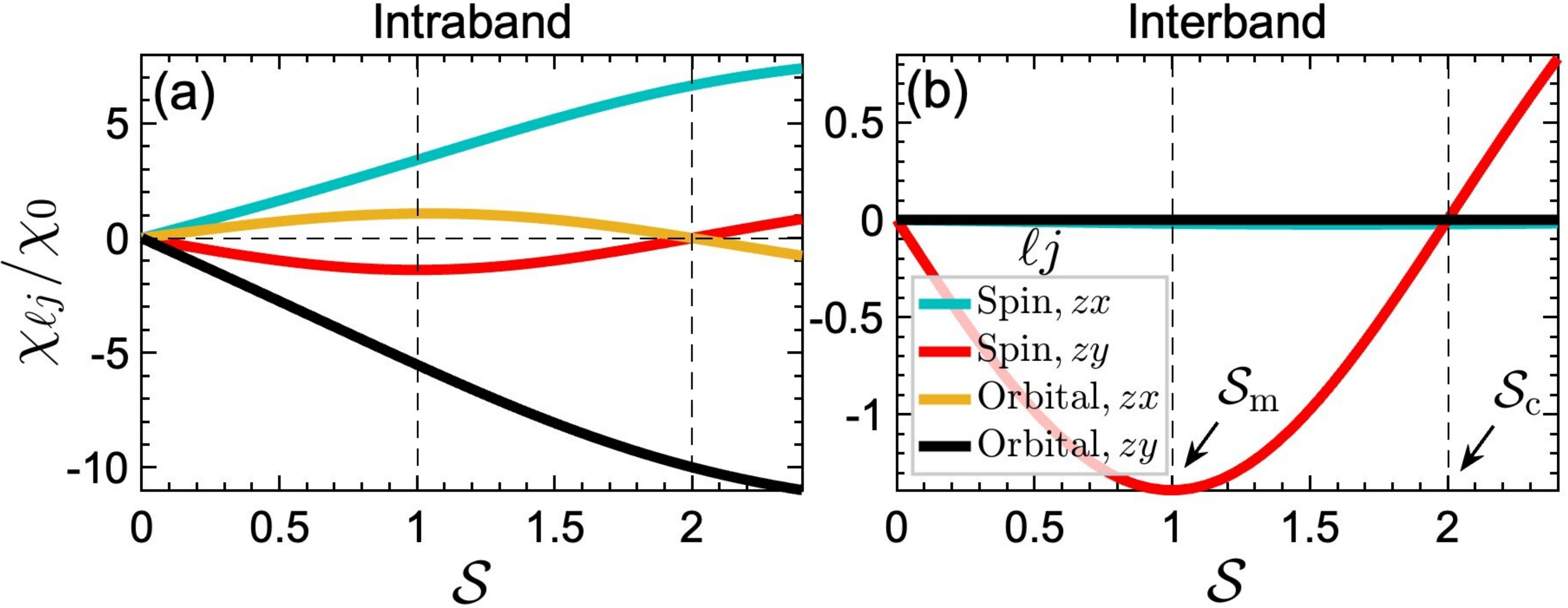}
			\caption{Intraband (a) and interband (b) contributions to the out-of-plane spin and orbital Edelstein susceptibilities $\chi_{zx}/\chi_0$ and $\chi_{zy}/\chi_0$ in a 2D Rashba $d_{x^2-y^2}$-wave AM as functions of the two-color circularly polarized parameter $\mathcal{S}$ at $e \mathcal{A}_0/\hbar k_{\rm F}=1$ and $\eta_1=\eta_2=+1$. In the intraband regime, all components vary smoothly with $\mathcal{S}$ and are dominated by the orbital channel, whereas in the interband regime only the spin-$zy$ component remains finite and exhibits a pronounced nonmonotonic dependence. Both channels show a sign reversal at the critical driving strength $\mathcal{S}_c=2$.}
			\label{f2}
		\end{figure} 
		
		In the intraband sector [Fig.~\ref{f2}(a)], the response depends strongly and monotonically on $\mathcal{S}$. Although the spin and orbital channels generally exhibit competing tendencies, the orbital contribution ultimately dominates. Its pronounced negative slope indicates that a stronger second-harmonic weight enhances the orbital Edelstein response by reshaping the band geometry and Berry curvature. In the interband sector [Fig.~\ref{f2}(b)], most components are suppressed in our continuum model (though transitions should emerge at higher energies in lattice models). However, the spin-$zy$ response dominates and scales strongly with $\mathcal{S}$, highlighting the pronounced sensitivity of the interband spin textures to the second driving field.
		
		Our dynamical scheme generates $\mathcal{C}_{2z}$-breaking in-plane Zeeman fields of $\mathcal{B} \approx 10$--$30~\mathrm{meV}$ (equivalent to $B \sim 85$--$250~\mathrm{T}$, estimated from $\mathcal{B} = g\mu_{\rm B} B$ with $g= 2$), bypassing the need for impractical static magnetic fields. Under two-color optical driving at $E_0 \sim 3$--$5~\mathrm{V/nm}$ ($\mathcal{S}=1$--$2$, $e\mathcal{A}_0/\hbar k_{\rm F}=1$--$2$), we consider typical $d$-wave UMs~\cite{Jiang2025,PhysRevX.12.031042,Weber2025,Reichlova2024,Feng2022,doi:10.1126/sciadv.aaz8809,PhysRevLett.128.197202,weber2024opticalexcitationspinpolarization} with a lattice constant $a \approx 3$~\AA, $k_{\rm F} \approx 0.1\pi/a$, and an exchange splitting $M_d=1 \approx 41.8$ meV. For these parameters, we estimate a robust and giant PEP ($O=S,L$)\begin{align}
			\langle O_z\rangle \simeq 0.5-1.5\,\mu_{\rm B}\,.
		\end{align}Such a massive out-of-plane accumulation is technologically vital, as it provides a highly efficient, field-free mechanism to exert torques and switch perpendicular magnetic memory devices~\cite{shao2021roadmap,Kang2025,Nguyen2024,clzw-1-2-022201,RevModPhys.91.035004}.
		
		The chiral two-color drive ($\eta_1,\eta_2$) defines a characteristic scale where the unidirectional ($zx$ or $zy$) polarization vanishes and changes sign. Importantly, the different polarization channels do not vanish simultaneously at this critical point. According to Eqs.~\eqref{eq_4b} and~\eqref{eq_4c}, the light-induced Rashba corrections selectively cancel terms proportional to $k_y$ and $k_x$. This yields $v_y = 0$ when $\lambda_{\rm R} = + \lambda_{\rm R\omega}$, extinguishing the spin $zy$ Edelstein component. Conversely, the Berry-curvature-driven orbital $zx$ component vanishes under the opposite condition, $\lambda_{\rm R} = - \lambda_{\rm R\omega}$. This degree-of-freedom-dependent zeroing yields the critical driving strength{\small\begin{align}
				\mathcal{S}^2_{\rm c} = \frac{2 \hbar \omega m_{\rm e}}{\eta_2 e^2 \mathcal{A}_0^2 M^{d}_{\rm C}} - \frac{2 \eta_1}{\eta_2}\,.
		\end{align}}For the parameters in Fig.~\ref{f2}, $\hbar \omega = 3~\text{eV}$, $M^{d}_{\rm C}=1$, $e \mathcal{A}_0=\hbar k_{\rm F}=1$, $m_{\rm e}=1$, and $\eta_1=\eta_2=+1$ (also for $\eta_1=\eta_2=-1$), this yields $\mathcal{S}_{\rm c}=2$. No critical point appears for $\eta_1=\mp \eta_2$. The sign reversal reflects competition between the fundamental and second-harmonic light-induced fields: the response is dominated by the fundamental for $\mathcal{S}<\mathcal{S}_{\rm c}$, vanishes at $\mathcal{S}_{\rm c}$ via destructive interference, and reverses for $\mathcal{S}>\mathcal{S}_{\rm c}$ as the second harmonic dominates, corresponding to a harmonic-controlled inversion of the nonequilibrium band geometry governing the Edelstein response.\begin{figure}[t]
			\centering
			\includegraphics[width=0.9\linewidth]{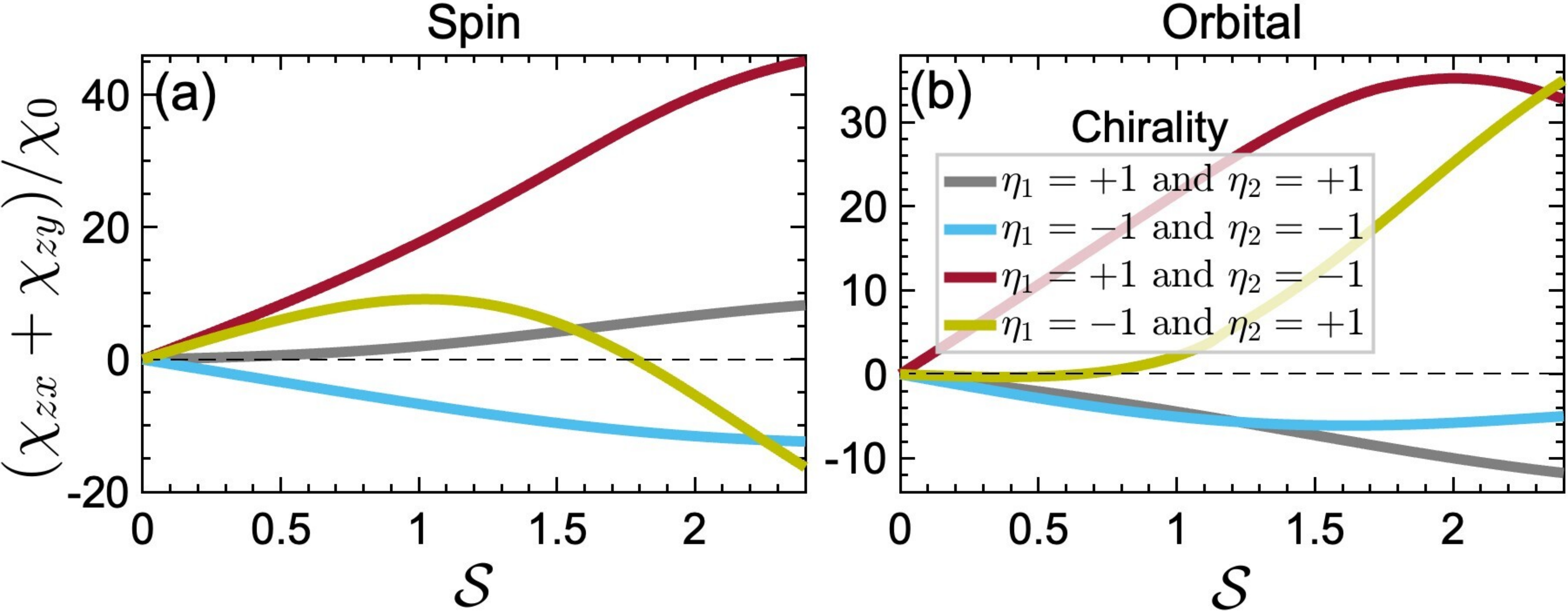}
			\caption{Spin (a) and orbital (b) contributions to the total out-of-plane Edelstein response $\chi_{zx}+\chi_{zy}$ as a function of $\mathcal{S}$ for different circular beam chiralities in a 2D Rashba $d_{x^2-y^2}$-wave AM. Here, $\eta_1$ and $\eta_2$ denote the chiralities of the first and second circular beams. For identical chiralities ($\eta_1=\eta_2=\pm1$) or opposite chiralities ($\eta_1=-\eta_2$) at $e \mathcal{A}_0/\hbar k_{\rm F}=1$, the out-of-plane Edelstein response is strongly modulated and can change sign.}
			\label{f3}
		\end{figure}
		
		Moreover, at $\mathcal{S}_{\rm m}=1$, where the spin-$zy$ component is minimized and the orbital-$zx$ component is maximized, the condition corresponds to a vanishing out-of-plane Hamiltonian term, $\mathcal{B}^d_z = - M_{\mathbf{k}}^{d}$. This yields $\mathcal{S}^2_{\rm m}
		=\frac{\hbar^5 \omega M_1^{d} (k_x^2 - k_y^2)}
		{\eta_2 e^2 \lambda_{\rm R}^2 \mathcal{A}_0^2 m_{\rm e}}
		- \frac{2\eta_1}{\eta_2}.$ For the chosen parameters, this gives $\mathcal{S}_{\rm m}=1$ provided $k_x^2 - k_y^2 = k_{\rm F}^2 \left(\frac{m_{\rm e} \lambda_{\rm R}^2}{\hbar^2 M_1^{d}}\right).$ Thus, by tuning the altermagnetic order relative to RSOC, the spin and orbital polarizations can be driven to their extrema.
		
		
		Next, in Fig.~\ref{f3}, we examine the role of the chirality of the two driving beams in generating out-of-plane polarizations. To avoid overcrowding, we present the total out-of-plane Edelstein response ($zx+zy$) assuming identical DC fields, $E_x=E_y$, rather than separate components. The response is highly sensitive to the relative chirality of the beams. For identical chiralities ($\eta_1=\eta_2=\pm1$), spin and orbital contributions vary smoothly and monotonically with $\mathcal{S}$ (gray and blue curves), reflecting constructive interference. In contrast, for opposite chiralities ($\eta_1=-\eta_2$), the response becomes strongly nonmonotonic and can reverse sign (red and green curves), indicating competition between contributions to the effective out-of-plane field. Nevertheless, both spin and orbital responses remain finite across all $\mathcal{S}$, demonstrating robust out-of-plane polarization despite partial cancellation effects.
		
		\blue{\textit{Experimental perspective}}---Our results indicate that PEPs in Rashba UMs are accessible under realistic conditions. Appropriate parameters match established $d$-wave altermagnets (e.g., RuO$_2$~\cite{Feng2022,doi:10.1126/sciadv.aaz8809,PhysRevLett.128.197202,weber2024opticalexcitationspinpolarization}, KV$_2$Se$_2$O~\cite{Jiang2025}, KRu$_4$O$_8$~\cite{PhysRevX.12.031042,Weber2025}, and Mn$_5$Si$_3$~\cite{Reichlova2024}), where the intersublattice spin texture is governed by an RSOC scale of $\lambda_{\rm R}\sim 0.3~\mathrm{eV}$\AA. To achieve the requisite Floquet field strengths without excessive heating, steady-state driving can be approximated using ultrashort, phase-locked $\omega$-$n\omega$ visible to near-UV pulses ($\hbar\omega\sim 1$--$5~\mathrm{eV}$). Under these conditions, the emergent spin textures are well within TR-ARPES resolution limits. Crucially, our bichromatic drive fundamentally modifies the system's Hamiltonian, thereby modulating the spin and orbital Edelstein effects. Although probing bare orbital currents remains challenging, electrically generated orbital accumulations are routinely resolved via ferromagnet torque manipulation and longitudinal Kerr/Faraday rotation~\cite{PhysRevLett.132.226704,PhysRevB.109.014420,PhysRevLett.132.236702,PhysRevMaterials.7.L111401,PhysRevLett.128.067201,PhysRevB.103.L020407,PhysRevLett.125.177201}. Consequently, these established techniques can readily detect our predicted light-induced responses. Ultimately, this framework offers a practical route to deterministic spintronic switching: transferring a transient angular momentum impulse to an adjacent perpendicular-anisotropy layer during the prethermal Floquet lifetime can seamlessly dictate post-pulse magnetic reversal.
		
		\blue{\textit{Conclusion}}---In summary, while nonrelativistic PEPs are permitted in equilibrium $p$-wave magnets, rigid spin-group symmetries strictly forbid these functional responses in materials featuring higher-order harmonics, such as $d$-wave altermagnets. We establish that two-color Floquet engineering, when hybridized with Rashba spin-orbit interactions, successfully bypasses these static spatial constraints to unlock massive PEPs in otherwise forbidden 2D spin-polarized magnets. Crucially, standard single-frequency optical drives are fundamentally incapable of lifting this restriction. By explicitly fracturing the $\mathcal{C}_{2z}$ rotational invariance, the nonlinear optical interference generates an emergent, planar Zeeman splitting without requiring external magnetic fields. 
		
		This dynamic symmetry manipulation drives robust out-of-plane accumulations where the orbital angular momentum fundamentally eclipses the competing spin channel. Because the direction and magnitude of these light-induced polarizations are deterministically dictated by the lattice's momentum-dependent magnetic parity and the chosen driving harmonics, our protocol ensures highly tunable, on-demand symmetry control. Ultimately, this framework provides a versatile, ultrafast pathway for the field-free perpendicular writing of memory devices in next-generation spintronics and orbitronics.
		
		\blue{\textit{Acknowledgments}}---M.\,Y. gratefully acknowledges useful discussions with Jairo Sinova and Rafael M. Fernandes and the hospitality of Uppsala University during his visit, where this work was performed. M.\,Y. and J.\,K.\,F. were supported by the Department of Energy, Office of Basic Energy Sciences, Division of Materials Sciences and Engineering under Contract No.\ DE-FG02-08ER46542 for the formal developments, the analytical/numerical work, and the writing of the manuscript. J.\,K.\,F. was also supported by the McDevitt bequest at Georgetown University. D.\,J., M.\,B.,
		and P.\,M.\,O. acknowledge support by the Swedish Research
		Council (VR), the German Research Foundation (Deutsche
		Forschungsgemeinschaft) through CRC/TRR 227 “Ultrafast
		Spin Dynamics” (project MF, Project ID 328545488), and
		the K. and A. Wallenberg Foundation (Grants No.\ 2022.0079 and No.\ 2023.0336). Part of the calculations were supported
		by the National Academic Infrastructure for Supercomputing in Sweden (NAISS) at NSC Link\"oping, funded by VR
		through Grant No.\ 2022-06725. L. Š.\ acknowledges funding from the ERC Starting Grant No.\ 101165122.
		
		\blue{\textit{Data availability}}---The data that support the findings of this article are openly available at~\cite{Zenodo}.
	}
	\bibliography{bib.bib}
	{\allowdisplaybreaks
		\onecolumngrid
		\subsection{\large End Matter}
		\twocolumngrid
		\hypertarget{mylinkA}{\blue{\textit{Derivation of the Effective Floquet Hamiltonian}}}---In the off-resonant high-frequency regime ($\hbar\omega$ well above the electronic bandwidth), the time-periodic Hamiltonian $H_{d}(\mathbf{k} + e\mathbf{A}(t)/\hbar)$ is treated with the van Vleck inverse-frequency expansion. The effective time-independent Hamiltonian up to order $1/\omega$ reads\begin{align}
			\mathcal{H}^{\rm eff}_{d}(\mathbf{k}) \approx {} &\mathcal{H}^{\rm F}_0(\mathbf{k}) + \sum_{m=1,2}\frac{1}{m\hbar\omega}\bigl[\mathcal{H}^{\rm F}_{-m}(\mathbf{k}),\,\mathcal{H}^{\rm F}_{+m}(\mathbf{k})\bigr],
		\end{align}where the Fourier components are
		$\mathcal{H}^{\rm F}_o(\mathbf{k}) = \frac{\omega}{2\pi}\int_0^{2\pi/\omega} dt\; H_{d}\bigl(\mathbf{k} + e\mathbf{A}(t)/\hbar\bigr)\, e^{io\omega t}$. Substituting the Peierls substitution and expanding to second order in $\mathbf{A}(t)$ yields{\small\begin{align}
				&H_{d}(\mathbf{k} + \mathbf{A}(t)) = H_{d}(\mathbf{k}) + \frac{e^2}{m_{\rm e}}M_2^{d}\sigma_z A_x(t)A_y(t)\notag \\ &
				+ \Bigl[\bigl(2\tfrac{e}{\hbar}\varepsilon\sigma_0 + \tfrac{e\hbar}{m_{\rm e}}M_1^{d}\sigma_z\bigr)k_x + \tfrac{e}{\hbar}\lambda_{\rm R}\sigma_y + \tfrac{e\hbar}{m_{\rm e}}M_2^{d}\sigma_z k_y\Bigr]A_x(t)\notag \\ &
				+ \Bigl[\bigl(2\tfrac{e}{\hbar}\varepsilon\sigma_0 - \tfrac{e\hbar}{m_{\rm e}}M_1^{d}\sigma_z\bigr)k_y - \tfrac{e}{\hbar}\lambda_{\rm R}\sigma_x + \tfrac{e\hbar}{m_{\rm e}}M_2^{d}\sigma_z k_x\Bigr]A_y(t)\notag \\ &
				+ \Bigl(\tfrac{e^2}{\hbar^2}\varepsilon\sigma_0 + \tfrac{e^2}{m_{\rm e}}\frac{M_1^{d}}{2}\sigma_z\Bigr)A_x^2(t) + \Bigl(\tfrac{e^2}{\hbar^2}\varepsilon\sigma_0 - \tfrac{e^2}{m_{\rm e}}\frac{M_1^{d}}{2}\sigma_z\Bigr)A_y^2(t).
		\end{align}}The Fourier integrals over one period $T = 2\pi/\omega$ for the relevant powers of $A_x(t)$ and $A_y(t)$ (and their products) are evaluated analytically. The results for $o = 0, \pm1, \pm2$ lead to the following structure:{\small\begin{align}\label{eq_12}
				\mathcal{H}^{\rm F}_o(\mathbf{k}) = h_0^{(o)}(\mathbf{k})\sigma_0 + h_x^{(o)}(\mathbf{k})\sigma_x + h_y^{(o)}(\mathbf{k})\sigma_y + h_z^{(o)}(\mathbf{k})\sigma_z,	\end{align}}with{\small\begin{subequations}\label{eq_11}
				\begin{align}
					&	h_0^{(0)}(\mathbf{k}) = {} \frac{\mathcal{A}_0^2 e^2 \varepsilon \left( 1 + S^2  \right)}{\hbar^2} ,\\
					&	h_{x,y,z}^{(0)}(\mathbf{k}) = {} 0\,,\\
					&	h_0^{(\pm 1)}(\mathbf{k}) = {}  \frac{2 \mathcal{A}_0 e \varepsilon \hbar \left(k_x\pm  i \eta_1 k_y\right) + \mathcal{A}_0^2 S e^2 \varepsilon e^{\mp i \phi} \left(  1 + \eta_1 \eta_2\right)}{2 \hbar^2}\, ,\\
					&	h_x^{(\pm 1)}(\mathbf{k}) = {}  \mp\frac{ i \mathcal{A}_0 e \eta_1 \lambda_{\rm R}}{2 \hbar}\, ,\\
					&	h_y^{(\pm 1)}(\mathbf{k}) = {}  \frac{\mathcal{A}_0 e \lambda_{\rm R}}{2 \hbar}
					\, ,\\
					&	h_z^{(\pm 1)}(\mathbf{k}) = {} \frac{\mathcal{A}_0 e}{4 m_{\rm e}} e^{\mp i\phi} \Bigl(
					2\hbar e^{\pm i\phi} (M_1^{d} k_x + M_2^{d} k_y) 
					\notag \\ {} &\mp  2i \eta_1 \hbar e^{\pm i\phi} (M_1^{d} k_y - M_2^{d} k_x) 
					\notag \\ {} &+ \mathcal{A}_0 \mathcal{S} e \Bigl[ 
					M_1^{d} (1 - \eta_1\eta_2) 
					\pm i M_2^{d} (\eta_2 - \eta_1)
					\Bigr]
					\Bigr)\,,\\
					&h_0^{(\pm 2)}(\mathbf{k}) = {}  \frac{\mathcal{A}_0 S e \varepsilon \left(k_x\pm i  \eta_2 k_y\right) e^{\mp i \phi}}{\hbar}
					\, ,\\
					&h_x^{(\pm 2)}(\mathbf{k}) = {} \mp  \frac{i  \mathcal{A}_0 S e \eta_2 \lambda_{\rm R}  \, e^{\mp i \phi}}{2 \hbar}
					\, ,\\
					&h_y^{(\pm 2)}(\mathbf{k}) = {}  \frac{e \lambda_{\rm R} \mathcal{A}_0 \mathcal{S}}{2 \hbar}e^{\mp i\phi} \, ,\\
					&h_z^{(\pm 2)}(\mathbf{k}) = {}  \frac{\mathcal{A}_0 e}{4 m_{\rm e}} e^{\mp i \phi} \bigg(\mathcal{A}_0  e e^{\pm i \phi} \left(M_1^{d} \pm i M_2^{d} \eta_1 \right)\notag \\ {}& + 2 \mathcal{S}\hbar \left(M_1^{d} k_x + M_2^{d} k_y \pm  i \eta_2\left(M_2^{d}k_x - M_1^{d} k_y\right)\right)\bigg)\, .
				\end{align}
		\end{subequations}}
		
		Evaluating the commutators generates a scalar shift, a renormalized SOC, and an effective Zeeman field. Since scalar components naturally commute, only the vector parts ($\mathbf{h}^{(o)} \cdot \boldsymbol{\sigma}$) contribute to these dynamical corrections. Applying the identity $[\mathbf{h}_1 \cdot \boldsymbol{\sigma},\, \mathbf{h}_2 \cdot \boldsymbol{\sigma}] = 2i\, (\mathbf{h}_1 \times \mathbf{h}_2) \cdot \boldsymbol{\sigma}$, the expansion terms yield corrections strictly proportional to the cross products $\mathbf{h}^{(-o)} \times \mathbf{h}^{(+o)}$. Collecting these single- and two-photon $\omega$--$2\omega$ interferences and adding them to the bare Hamiltonian yields the final effective model.
		
		\hypertarget{mylinkB}{\blue{\textit{The effective Floquet Hamiltonian for two linearly and hybrid linearly-circularly polarized laser fields}}}---For two linearly polarized laser fields, the vector potential reads $A_x(t) = {} \mathcal{A}_0\bigl[\cos(\omega t) + \mathcal{S}\cos(2\omega t + \phi) \cos\psi\bigr]$ and $A_y(t) = {} \mathcal{A}_0\mathcal{S}\cos(2\omega t + \phi) \sin\psi$, where $\psi$ is the polarization
		direction of the linearly polarized light. This leads to the effective Hamiltonian $\mathcal{H}^{\rm eff}_{d}(\mathbf{k}) ={} h^d_0(\mathbf{k})\sigma_0 + \textbf{h}^d(\mathbf{k})\cdot \boldsymbol{\sigma}$ with{\small\begin{subequations}
				\begin{align}
					h^d_0(\mathbf{k}) ={} & \varepsilon(k_x^2 + k_y^2) +  \frac{e^2}{2\hbar^2} \varepsilon \mathcal{A}_0^2 (1+ \mathcal{S}^2) \,, \\
					h^d_x(\mathbf{k}) = {} &- \lambda_{\rm R} k_y +\frac{e^3 \lambda_{\rm R} \mathcal{A}_0^3  \mathcal{S} \sin\phi}{8\hbar^2 m_{\rm e} \omega}
					\big(3 M_1^{d} \cos\psi + 4 M_2^{d} \sin\psi \big) \, ,\\
					h^d_y(\mathbf{k}) = {} &\lambda_{\rm R} k_x  -\frac{e^3 \lambda_{\rm R} \mathcal{A}_0^3  \mathcal{S}   \sin\phi   }{8\hbar^2 m_{\rm e} \omega} M_1^{d}\sin\psi \, ,\\
					h^d_z(\mathbf{k}) = {} &\frac{\hbar^2}{m_{\rm e}}\left(\tfrac{k_x^2 - k_y^2}{2}M_1^{d}  + k_x k_y M_2^{d}\right) \notag \\ {} &+\frac{e^2 \mathcal{A}_0^2  }{4 m_{\rm e}}
					\Big[ M_1^{d} +  \mathcal{S}^2 \left(M_1^{d} \cos (2\psi) + M_2^{d} \sin (2\psi)\right)  \Big]\, ,
				\end{align}
		\end{subequations}}
		
		Moreover, for two linearly-circularly polarized laser fields, the vector potential reads $A_x(t) = {} \mathcal{A}_0\bigl[\cos(\omega t) + \mathcal{S}\cos(2\omega t + \phi) \cos\psi\bigr]$ and $A_y(t) = {} \mathcal{A}_0\bigl[\eta_1 \sin(\omega t) + \mathcal{S}\cos(2\omega t + \phi) \sin\psi\bigr]$, leading to the effective Hamiltonian $\mathcal{H}^{\rm eff}_{d}(\mathbf{k}) ={} h^d_0(\mathbf{k})\sigma_0 + \textbf{h}^d(\mathbf{k})\cdot \boldsymbol{\sigma}$ with{\small\begin{subequations}
				\begin{align}
					&	h^d_0(\mathbf{k}) = {}  \varepsilon(k_x^2 + k_y^2) +  \frac{e^2}{2\hbar^2} \varepsilon \mathcal{A}_0^2 (2+ \mathcal{S}^2)\, ,\\
					&	h^d_x(\mathbf{k}) = {}  - \lambda_{\rm R} k_y + \frac{ e^2  \lambda_{\rm R} \mathcal{A}_0^2}{4\hbar^2 m_{\rm e} \omega}
					\Big[ 4 \eta_1 \hbar (M_1^{d} k_y - M_2^{d} k_x) \notag \\ {} &\hspace*{1cm} + \mathcal{A}_0 \mathcal{S} e \Big(
					M_1^{d} \cos\psi \sin\phi + 2 M_2^{d} \sin\psi \sin\phi \notag \\ {} &\hspace*{1cm}+ \eta_1 \big( M_2^{d} \cos\psi \cos\phi - 2 M_1^{d} \sin\psi \cos\phi \big)
					\Big)
					\Big]\,, \\
					&	h^d_y(\mathbf{k}) = {}  \lambda_{\rm R} k_x + \frac{e^2 \lambda_{\rm R} \mathcal{A}_0^2}{4\hbar^2 m_{\rm e} \omega}
					\Big[ 4  \eta_1 \hbar (M_1^{d} k_x + M_2^{d} k_y) \notag \\ {} &\hspace*{1cm}+ \mathcal{A}_0 \mathcal{S} e \Big(
					-2 M_2^{d} \cos\psi \sin\phi + M_1^{d} \sin\psi \sin\phi \notag \\ {} &\hspace*{1cm}+ \eta_1 \big( 2 M_1^{d} \cos\psi \cos\phi + M_2^{d} \sin\psi \cos\phi \big)
					\Big)
					\Big]\,, \\
					&	h^d_z(\mathbf{k}) = {}  \frac{\hbar^2}{m_{\rm e}}\left[\frac{k_x^2 - k_y^2}{2}M_1^{d}  + k_x k_y M_2^{d}\right]  \notag \\ {} &\hspace*{0.3cm}+ \frac{e^2 \mathcal{A}_0^2}{4 m_{\rm e}}
					\Big[  - \frac{4  \eta_1 \lambda_{\rm R}^2 m_{\rm e}}{\hbar^3 \omega} + \mathcal{S}^2\big( M_1^{d} \cos (2\psi) + M_2^{d} \sin(2\psi) \big)
					\Big]\, .
				\end{align}
		\end{subequations}}\begin{figure}[b]
			\centering
			\includegraphics[width=0.9\linewidth]{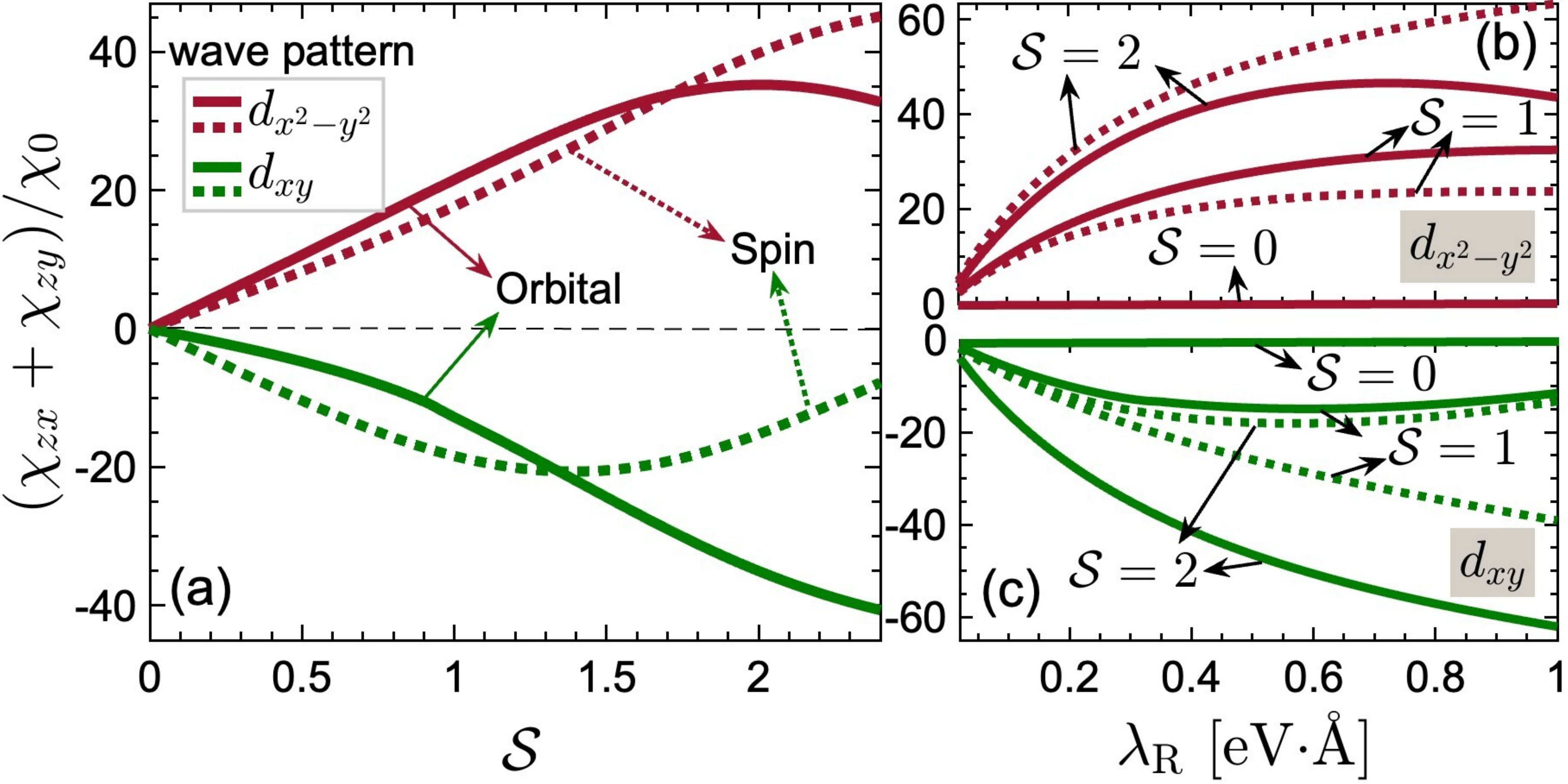}
			\caption{(a) Total out-of-plane Edelstein response $(\chi_{zx}+\chi_{zy})/\chi_0$ as a function of $\mathcal{S}$ for different symmetry wave patterns, $d_{x^2-y^2}$ and $d_{xy}$. Solid and dashed lines represent the spin and orbital contributions, respectively. (b,c) Dependence of the same response on the RSOC strength $\lambda_{\rm R}$ for selected values of $\mathcal{S}$. Panel (b) corresponds to the $d_{x^2-y^2}$ pattern, while panel (c) shows the $d_{xy}$ pattern. The parameters are $e \mathcal{A}_0/\hbar k_{\rm F} = 1$, $M_{d} = 1$, $\hbar \omega = 3~\mathrm{eV}$, $\eta_1 = +1$, and $\eta_2 = -1$. In (a), we fix $\lambda_{\rm R} = 0.3~\mathrm{eV}\cdot$\AA.}
			\label{f4}
		\end{figure}
		
		Notably, bi-linear and mixed linearly-circularly polarized drives also generate an analogous in-plane Zeeman-like field $\boldsymbol{\mathcal{B}}_\parallel$. While sharing the same symmetry-breaking mechanism, their distinct microscopic expressions yield quantitatively and qualitatively different out-of-plane Edelstein polarizations compared to the bi-circular case.
		
		\hypertarget{mylinkC}{\blue{\textit{Perpendicular Edelstein polarization for $d_{xy}$-wave AM and various RSOC strengths}}}---In Fig.~\ref{f4}, we consider the chirality configuration $\eta_1 = +1$ and $\eta_2 = -1$ from Fig.~\ref{f3}, which yields the strongest response, and examine how the $d$-wave AM patterns, together with the RSOC, influence the induced Edelstein polarizations. In Fig.~\ref{f4}(a), the response shows a pronounced dependence on the underlying AM wave pattern. For the $d_{x^2-y^2}$ pattern, both spin and orbital contributions are positive and increase with $\mathcal{S}$, indicating a cooperative enhancement of the out-of-plane polarization. Conversely, for the $d_{xy}$ pattern, both contributions are negative and scale with $\mathcal{S}$, reversing the polarization relative to the $d_{x^2-y^2}$ case. Across these distinct symmetries, a clear physical picture emerges: although spin and orbital polarizations frequently compete, the orbital channel fundamentally dominates the overall response, especially at larger $\mathcal{S}$. Figures~\ref{f4}(b) and~\ref{f4}(c) reveal how the response evolves with the RSOC $\lambda_{\rm R}$. For the $d_{x^2-y^2}$ pattern (b), the response increases and saturates at large $\lambda_{\rm R}$, amplified by higher $\mathcal{S}$. Conversely, the $d_{xy}$ channel (c) grows increasingly negative with $\lambda_{\rm R}$, though larger $\mathcal{S}$ still enhances its magnitude. Thus, the specific wave pattern and RSOC jointly dictate both the sign and magnitude of the induced out-of-plane polarization.

	}

	\onecolumngrid
	\clearpage
	
	{\allowdisplaybreaks
		
		\begin{center}
			\textbf{\large \vskip0mm Supplemental Materials for ``Giant perpendicular Edelstein polarizations in 2D compensated magnets via bichromatic Floquet driving''}
			\vskip3.5mm
			Mohsen Yarmohammadi,$^1$ Daegeun Jo,$^2$ Marco Berritta,$^2$ Libor \v{S}mejkal,$^{3,4,5}$ James K. Freericks,$^1$ and Peter M. Oppeneer$^2$\vskip1mm
			\small $^1$\textit{Department of Physics, Georgetown University, Washington DC 20057, USA}\\
			$^2$\textit{Department of Physics and Astronomy, Uppsala University, P.O. Box 516, SE-75120 Uppsala, Sweden}\\
			\small $^3$\textit{Max Planck Institute for the Physics of Complex Systems, N\"othnitzer Str.\ 38, 01187 Dresden, Germany}\\
			\small $^4$\textit{Max Planck Institute for Chemical Physics of Solids, N\"othnitzer Str.\ 40, 01187 Dresden, Germany}\\
			\small $^5$\textit{Institute of Physics, Czech Academy of Sciences, Cukrovarnick\'a 10, 162 00 Praha 6, Czech Republic}\\
			(Dated: \today)
		\end{center}
		\setcounter{equation}{0}
		\makeatletter

		\setcounter{equation}{0}
		\renewcommand{\theequation}{S\arabic{equation}}
		\setcounter{figure}{0}
		\renewcommand{\thefigure}{S\arabic{figure}}
		\setcounter{section}{0}
		\renewcommand{\thesection}{S\arabic{section}}
		\setcounter{table}{0}
		\renewcommand{\thetable}{S\arabic{table}}
		
		\section{S1. Effective Hamiltonian of 2D Rashba $\ell$-wave ($\ell=p,d,f,g$) magnets driven by two-color $\omega$--$2\omega$ inearly polarized light fields}
		In this section, we derive the effective Floquet Hamiltonians for a two-band system with various altermagnetic (AM) symmetries driven by a two-color linearly polarized light field. The pristine static Hamiltonian is given by
		\begin{equation}
			H_{\rm d}(\mathbf{k}) = \varepsilon(k_x^2 + k_y^2)\sigma_0 + \lambda_{\rm R}(k_x\sigma_y - k_y\sigma_x) + M_{\mathbf{k}}^{\rm d}\sigma_z,
		\end{equation}
		where $\varepsilon = \hbar^2/2m_e$, $\lambda_{\rm R}$ is the Rashba spin-orbit coupling, and $M_{\mathbf{k}}^{\rm d}$ dictates the altermagnetic wave symmetry.
		
		The external optical field is defined by the vector potentials
		\begin{subequations}
			\begin{align}
				A_x(t) &= \mathcal{A}_0 \big[ \cos(\omega t) + \mathcal{S}\cos(2\omega t + \psi)\cos\psi \big], \\
				A_y(t) &= \mathcal{S} \mathcal{A}_0 \cos(2\omega t + \psi)\sin\psi,
			\end{align}
		\end{subequations}
		where $\omega$ is the fundamental driving frequency, $ \mathcal{S}$ controls the relative amplitude of the second harmonic, $\psi$ is the phase delay, and $\psi$ sets the polarization angle.
		
		To couple the light to the electrons, we apply the minimal coupling formalism (Peierls substitution), shifting the crystalline momentum $\mathbf{k} \rightarrow \mathbf{k}(t)$ via
		\begin{equation}
			k_x(t) = k_x + \frac{e A_x(t)}{\hbar}, \quad k_y(t) = k_y + \frac{e A_y(t)}{\hbar}.
		\end{equation}
		Substituting $\mathbf{k}(t)$ into $H_{\rm d}(\mathbf{k})$ yields a time-dependent Hamiltonian $\mathcal{H}(\mathbf{k}, t)$. In the high-frequency limit ($\hbar\omega \gg \varepsilon, \lambda_{\rm R}, M$), the effective Hamiltonian $\mathcal{H}^{\rm eff}_{\rm d}(\mathbf{k})$ is obtained via the Magnus expansion by time-averaging the static terms and incorporating the leading-order dynamical corrections from the commutators of the oscillating terms (see End Matter in the main text for details of the Floquet-Magnus expansion):
		\begin{equation}
			\mathcal{H}^{\rm eff}_{\rm d}(\mathbf{k}) = h_0(\mathbf{k})\sigma_0 + \mathbf{h}(\mathbf{k})\cdot \boldsymbol{\sigma}.
		\end{equation}
		Below, in addition to the $d$-wave Rashba AMs in the main text, we present the components of $\mathcal{H}^{\rm eff}_{\rm d}(\mathbf{k})$ for $s$-, $p$-, $f$-, and $g$-wave AMs.
		
		\subsection*{S1.1. Isotropic Ponderomotive shift (common term)}
		
		For all AM symmetries considered here, the identity-matrix coefficient $h_0(\mathbf{k})$ collects the unperturbed kinetic energy and the time-averaged ponderomotive energy. It evaluates to:
		\begin{equation}
			h_0(\mathbf{k}) = \varepsilon(k_x^2 + k_y^2) + \frac{\varepsilon e^2 \mathcal{A}_0^2}{2\hbar^2}(1 + \mathcal{S}^2).
		\end{equation}
		
		\subsection*{S1.2. s-wave and p-wave magnets}
		
		For an $s$-wave order parameter $M_{\mathbf{k}}^{\rm s} = M \hbar^2/m_e$, the light field does not induce any higher-order momentum couplings in the exchange field. The components are simply the unperturbed Rashba and exchange terms, given by
		\begin{equation}
			h_x^{(s)} = -\lambda_{\rm R} k_y, \quad h_y^{(s)} = \lambda_{\rm R} k_x, \quad h_z^{(s)} = M \frac{\hbar^2}{m_e}.
		\end{equation}
		For a generic $p$-wave state ($M_{\mathbf{k}}^{\rm p} \propto M_C k_x + M_S k_y$), the leading-order time-averaged $h_z$ identically matches the unperturbed state since terms linear in $\mathbf{A}(t)$ time-average to zero. There are no in-plane $\textbf{k}$-independent Zeeman-like fields $\mathcal{B}_x$ and $\mathcal{B}_y$ present to break the $\mathcal{C}_2$ rotational symmetry.

		\subsection*{S1.3. f-wave magnets}
		
		For an $f$-wave AM state, the unperturbed term is $M_{\mathbf{k}}^{\rm f} = \frac{\hbar^2}{m_e} \big[ M_C k_x(k_x^2 - 3k_y^2) + M_S k_y(k_y^2 - 3k_x^2) \big]$. The effective Hamiltonian components read as
		\begin{subequations}
			\begin{align}
				h_x^{(f)} &= -\lambda_{\rm R} k_y + \frac{\lambda_{\rm R} \mathcal{A}_0^3 \mathcal{S} e^3 \sin\psi}{4\hbar^2 m_e \omega} \Big[ 9 M_C k_x \cos\psi - 9 M_S k_y \cos\psi - 12 M_C k_y \sin\psi - 12 M_S k_x \sin\psi \Big], \\
				h_y^{(f)} &= \lambda_{\rm R} k_x - \frac{3 \lambda_{\rm R} \mathcal{A}_0^3 \mathcal{S} e^3 \sin\psi \sin\psi}{4\hbar^2 m_e \omega} \Big[ M_C k_x - M_S k_y \Big], \\
				h_z^{(f)} &= M_{\mathbf{k}}^{\rm f} + \frac{3 \mathcal{A}_0^2 e^2}{2 m_e} \Big[ (1 + \mathcal{S}^2 \cos 2\psi)(M_C k_x - M_S k_y) - \mathcal{S}^2 \sin 2\psi (M_C k_y + M_S k_x) \Big] \\
				&\quad + \frac{3 \mathcal{A}_0^3 \mathcal{S} e^3 \cos\psi}{4 \hbar m_e} (M_C \cos\psi - M_S \sin\psi).
			\end{align}
		\end{subequations}In this case, in-plane $\textbf{k}$-independent Zeeman-like fields $\mathcal{B}_x$ and $\mathcal{B}_y$ are also absent.
		
		\subsection*{S1.4. g-wave magnets}
		
		The pristine $g$-wave term is $M_{\mathbf{k}}^{\rm g} = \frac{\hbar^2}{m_e} \big[ M_C(k_x^4 - 6k_x^2 k_y^2 + k_y^4) + 4M_S k_x k_y(k_x^2 - k_y^2) \big]$. To ensure readability, we group the light-induced corrections by powers of $\mathcal{A}_0$ as
		\begin{equation}
			h_x^{(g)} = -\lambda_{\rm R} k_y + \frac{\lambda_{\rm R}}{8\hbar^4 m_e \omega} \Big( \Delta h_{x, 3}^{(g)} + \Delta h_{x, 5}^{(g)} \Big),
		\end{equation}
		with
		\begin{subequations}
			\begin{align}
				\Delta h_{x, 3}^{(g)} &= 12 \mathcal{A}_0^3 \mathcal{S} e^3 \hbar^2 \sin\psi \Big[ 3 M_C \cos\psi (k_x^2 - k_y^2) + 4 M_S \sin\psi (k_x^2 - k_y^2) + 6 M_S k_x k_y \cos\psi - 8 M_C k_x k_y \sin\psi \Big], \\
				\Delta h_{x, 5}^{(g)} &= \mathcal{A}_0^5 \mathcal{S} e^5 \sin\psi \Big[ 6 M_C \cos\psi + 8 M_S \sin\psi - 33 M_C \mathcal{S}^2 \cos\psi - 12 M_S \mathcal{S}^2 \sin\psi \notag\\
				&\quad + 42 M_C \mathcal{S}^2 \cos^3\psi + 42 M_S \mathcal{S}^2 \cos^2\psi \sin\psi \Big].
			\end{align}
		\end{subequations}The $y$-component is also written as
		\begin{equation}
			h_y^{(g)} = \lambda_{\rm R} k_x - \frac{\lambda_{\rm R}}{8\hbar^4 m_e \omega} \Big( \Delta h_{y, 3}^{(g)} + \Delta h_{y, 5}^{(g)} \Big),
		\end{equation}
		with
		\begin{subequations}
			\begin{align}
				\Delta h_{y, 3}^{(g)} &= 12 \mathcal{A}_0^3 \mathcal{S} e^3 \hbar^2 \sin\psi \sin\psi \Big[ M_C (k_x^2 - k_y^2) + 2 M_S k_x k_y \Big], \\
				\Delta h_{y, 5}^{(g)} &= \mathcal{A}_0^5 \mathcal{S} e^5 \sin\psi \Big[ 2 M_C \sin\psi + 6 M_S \mathcal{S}^2 \cos\psi - 3 M_C \mathcal{S}^2 \sin\psi - 6 M_S \mathcal{S}^2 \cos^3\psi + 6 M_C \mathcal{S}^2 \cos^2\psi \sin\psi \Big].
			\end{align}
		\end{subequations}
		The $z$-component incorporates field-induced corrections at orders $\mathcal{O}(\mathcal{A}_0^2)$, $\mathcal{O}(\mathcal{A}_0^3)$, and $\mathcal{O}(\mathcal{A}_0^4)$:
		\begin{equation}
			h_z^{(g)} = M_{\mathbf{k}}^{\rm g} + \frac{1}{8\hbar^2 m_e} \Big( \Delta h_{z, 2}^{(g)} + \Delta h_{z, 3}^{(g)} + \Delta h_{z, 4}^{(g)} \Big),
		\end{equation}
		where
		\begin{subequations}
			\begin{align}
				\Delta h_{z, 2}^{(g)} &= \mathcal{A}_0^2 e^2 \hbar^2 \Big[ 24 M_C (k_x^2 - k_y^2) + 48 M_S k_x k_y + \mathcal{S}^2 \big( 4 M_C + 48 M_C k_x^2 \cos^2\psi - 24 M_C k_y^2 \cos^2\psi\notag \\
				&\quad - 24 M_C k_x^2 - 4 M_C \cos^2\psi + 24 M_S k_x^2 \sin(2\psi) - 24 M_S k_y^2 \sin(2\psi) \notag\\
				&\quad + 96 M_S k_x k_y \cos^2\psi - 48 M_S k_x k_y - 48 M_C k_x k_y \sin(2\psi) \big) \Big], \\
				\Delta h_{z, 3}^{(g)} &= 24 \mathcal{A}_0^3 \mathcal{S} e^3 \hbar \cos\psi \Big[ M_C(k_x \cos\psi - k_y \sin\psi) + M_S(k_y \cos\psi + k_x \sin\psi) \Big], \\
				\Delta h_{z, 4}^{(g)} &= \mathcal{A}_0^4 e^4 \Big[ 3 M_C - 12 M_C \mathcal{S}^2 + 24 M_C \mathcal{S}^2 \cos^2\psi - 18 M_C \mathcal{S}^4 \cos^2\psi + 21 M_C \mathcal{S}^4 \cos^4\psi \notag\\
				&\quad + 12 M_S \mathcal{S}^2 \sin(2\psi) - 6 M_S \mathcal{S}^4 \sin(2\psi) + 24 M_S \mathcal{S}^4 \cos^3\psi \sin\psi \Big].
			\end{align}
		\end{subequations}In this case, $\Delta h_{x/y, 5}^{(g)}$ are the in-plane $\textbf{k}$-independent Zeeman-like fields, which break the $\mathcal{C}_2$ rotational symmetry.

		\section{S2. Effective Hamiltonian of 2D Rashba $\ell$-wave ($\ell=p,d,f,g$) magnets driven by two-color $\omega$--$2\omega$ circularly polarized light fields}
		In this section, we derive the effective Floquet Hamiltonians for a two-band system with various AM symmetries driven by a two-color circularly polarized light field. The external optical field is defined by the vector potentials
		\begin{subequations}
			\begin{align}
				A_x(t) &= \mathcal{A}_0 \big[ \cos(\omega t) + \mathcal{S}\cos(2\omega t + \psi) \big], \\
				A_y(t) &= \mathcal{A}_0 \big[ \eta_1 \sin(\omega t) + \eta_2 \mathcal{S} \sin(2\omega t + \psi) \big],
			\end{align}
		\end{subequations}
		where $\omega$ is the fundamental driving frequency, $\mathcal{S}$ controls the relative amplitude of the second harmonic, $\psi$ is the phase delay, and $\eta_1, \eta_2 \in \{\pm 1\}$ determine the specific helicities of the fundamental and second-harmonic fields, respectively. Below, we detail the effective components for $s$-, $p$-, $f$-, $g$-, and $i$-wave AMs.
		
		\subsection*{S2.1. Isotropic Ponderomotive shift (common term)}
		
		Similarly, for all AM symmetries considered, the trace coefficient $h_0(\mathbf{k})$ consists of the pristine kinetic energy modified by the time-averaged ponderomotive energy
		\begin{equation}
			h_0(\mathbf{k}) = \varepsilon(k_x^2 + k_y^2) + \frac{\varepsilon e^2 \mathcal{A}_0^2}{\hbar^2}(1 + \mathcal{S}^2).
		\end{equation}
		
		\subsection*{S2.2. s-wave magnets}
		
		For an $s$-wave order parameter $M_{\mathbf{k}}^{\rm s} = M \hbar^2/m_e$, the effective components are
		\begin{subequations}
			\begin{align}
				h_x^{(s)} &= -\lambda_{\rm R} k_y, \\
				h_y^{(s)} &= \lambda_{\rm R} k_x, \\
				h_z^{(s)} &= M \frac{\hbar^2}{m_e} - \frac{\mathcal{A}_0^2 e^2 \lambda_{\rm R}^2}{2\hbar^3 m_e \omega} \big( 2\eta_1 + \eta_2 \mathcal{S}^2 \big).
			\end{align}
		\end{subequations}
		The additional term in $h_z$ represents the canonical effective Zeeman field induced by the circularly polarized light acting on the Rashba bands. Similar to the two-color linearly polarized light, there are no in-plane $\textbf{k}$-independent Zeeman-like fields $\mathcal{B}_x$ and $\mathcal{B}_y$ present to break the $\mathcal{C}_2$ rotational symmetry.
		
		\subsection*{S2.3. p-wave magnets}
		
		For a $p$-wave state $M_{\mathbf{k}}^{\rm p} = \frac{\hbar^2}{m_e}(M_C k_x + M_S k_y)$, the components are
		\begin{subequations}
			\begin{align}
				h_x^{(p)} &= -\lambda_{\rm R} k_y - \frac{\lambda_{\rm R} \mathcal{A}_0^2 e^2 M_S}{2\hbar m_e \omega} \big( 2\eta_1 + \eta_2 \mathcal{S}^2 \big), \\
				h_y^{(p)} &= \lambda_{\rm R} k_x + \frac{\lambda_{\rm R} \mathcal{A}_0^2 e^2 M_C}{2\hbar m_e \omega} \big( 2\eta_1 + \eta_2 \mathcal{S}^2 \big), \\
				h_z^{(p)} &= M_{\mathbf{k}}^{\rm p} - \frac{\mathcal{A}_0^2 e^2 \lambda_{\rm R}^2}{2\hbar^3 m_e \omega} \big( 2\eta_1 + \eta_2 \mathcal{S}^2 \big).
			\end{align}
		\end{subequations}
		Here, the light field generates constant momentum-independent shifts in the in-plane pseudo-spin components $h_{x,y}$, indicative of an effective light-induced persistent current. Thus, in-plane $\textbf{k}$-independent Zeeman-like fields $\mathcal{B}_x$ and $\mathcal{B}_y$ are present.
		
		\subsection*{S2.4. f-wave magnets}
		
		For an $f$-wave AM state, $M_{\mathbf{k}}^{\rm f} = \frac{\hbar^2}{m_e} \big[ M_C k_x(k_x^2 - 3k_y^2) + M_S k_y(k_y^2 - 3k_x^2) \big]$. The effective components expand as
		\begin{subequations}
			\begin{align}
				h_x^{(f)} &= -\lambda_{\rm R} k_y + \frac{3 \lambda_{\rm R} \mathcal{A}_0^2 e^2}{4\hbar m_e \omega} \Delta h_{x, 2}^{(f)} + \frac{\lambda_{\rm R} \mathcal{A}_0^4 e^4}{8\hbar^3 m_e \omega} \Delta h_{x, 4}^{(f)} + \frac{3 \lambda_{\rm R} \mathcal{A}_0^3 \mathcal{S} e^3}{4\hbar^2 m_e \omega} \Delta h_{x, 3}^{(f)}, \\
				h_y^{(f)} &= \lambda_{\rm R} k_x - \frac{3 \lambda_{\rm R} \mathcal{A}_0^2 e^2}{4\hbar m_e \omega} \Delta h_{y, 2}^{(f)} - \frac{\lambda_{\rm R} \mathcal{A}_0^4 e^4}{8\hbar^3 m_e \omega} \Delta h_{y, 4}^{(f)} + \frac{3 \lambda_{\rm R} \mathcal{A}_0^3 \mathcal{S} e^3}{4\hbar^2 m_e \omega} \Delta h_{y, 3}^{(f)}, \\
				h_z^{(f)} &= M_{\mathbf{k}}^{\rm f} + \frac{\mathcal{A}_0^2 e^2 \lambda_{\rm R}^2}{2\hbar^3 m_e \omega} \big( 2\eta_1 + \eta_2 \mathcal{S}^2 \big) - \frac{3 \mathcal{A}_0^3 \mathcal{S} e^3}{2 \hbar m_e} \Big[ M_C(\eta_1 \eta_2 + 1)\cos\psi + M_S(\eta_2 - \eta_1)\sin\psi \Big].
			\end{align}
		\end{subequations}
		where the shift functions are
		\begin{subequations}
			\begin{align}
				\Delta h_{x, 2}^{(f)} &= M_S \Big[ \eta_1(k_x^2 - k_y^2) + \frac{\eta_2 \mathcal{S}^2}{2}(k_x^2 - k_y^2) \Big] + 2 M_C k_x k_y \Big[ \eta_1 + \frac{\eta_2 \mathcal{S}^2}{2} \Big], \\
				\Delta h_{x, 4}^{(f)} &= \frac{3}{4}M_S \Big[ 2\eta_1(1 - \eta_1^2) + \mathcal{S}^4 \eta_2 (1 - \eta_2^2) \Big] = 0 \quad (\text{since } \eta_i^2 = 1), \\
				\Delta h_{x, 3}^{(f)} &= \sin\psi \Big[ M_C k_x(1 - \eta_1 \eta_2) - M_S k_y(1 + \eta_1 \eta_2) \Big] - \cos\psi \Big[ M_S k_x(\eta_1 - \eta_2) + M_C k_y(\eta_1 + \eta_2) \Big], \\
				\Delta h_{y, 2}^{(f)} &= M_C \Big[ \eta_1(k_x^2 - k_y^2) + \frac{\eta_2 \mathcal{S}^2}{2}(k_x^2 - k_y^2) \Big] - 2 M_S k_x k_y \Big[ \eta_1 + \frac{\eta_2 \mathcal{S}^2}{2} \Big], \\
				\Delta h_{y, 3}^{(f)} &= \sin\psi \Big[ M_S k_x(1 + \eta_1 \eta_2) - M_C k_y(1 - \eta_1 \eta_2) \Big] + \cos\psi \Big[ M_C k_x(\eta_1 + \eta_2) - M_S k_y(\eta_1 - \eta_2) \Big].
			\end{align}
		\end{subequations}In this case, in-plane $\textbf{k}$-independent Zeeman-like fields $\mathcal{B}_x$ and $\mathcal{B}_y$ are absent.
		
		\subsection*{S2.5. g-wave magnets}
		
		The pristine $g$-wave term is $M_{\mathbf{k}}^{\rm g} = \frac{\hbar^2}{m_e} \big[ M_C(k_x^4 - 6k_x^2 k_y^2 + k_y^4) + 4M_S k_x k_y(k_x^2 - k_y^2) \big]$. Grouping the corrections
		\begin{equation}
			h_x^{(g)} = -\lambda_{\rm R} k_y + \Delta h_{x}^{(g)}(\textbf{k}) +  \Delta h_{x}^{(g)}(\textbf{k} = 0),
		\end{equation}where $\Delta h_{x}^{(g)}(\mathbf{k})$ is lengthy and is not explicitly shown here. However, for our purposes, it is important to note that there exists a $\mathbf{k}$-independent contribution giving rise to $\mathcal{B}_x$, given by
		\begin{align}
			\Delta h_x^{(g)}(\mathbf{k}=0) ={} & \frac{\lambda_{\rm R} \mathcal{A}_0^3 \mathcal{S} e^3}{16 \hbar^4 m_e \omega} \Bigg[ 4 \hbar^2 M_C (1 + 4\eta_1\eta_2) \sin\psi 
			\notag \\ {} &	+ 2 \mathcal{A}_0^2 e^2 \Big( 4 M_S (\eta_1 - \eta_2)(3\mathcal{S}^2 - 4) \cos\psi + 3 M_C \big[ \mathcal{S}^2(1 - 8\eta_1\eta_2) - 6 \big] \sin\psi \Big) \Bigg],
		\end{align}
		The $y$-component follows structurally similarly as
		\begin{equation}
			h_y^{(g)} = \lambda_{\rm R} k_x + \Delta h_{y}^{(g)}(\textbf{k}) +  \Delta h_{y}^{(g)}(\textbf{k} = 0),
		\end{equation}with
		\begin{align}
			\Delta h_y^{(g)}(\mathbf{k}=0) = {} &\frac{\lambda_{\rm R} \mathcal{A}_0^3 \mathcal{S} e^3}{8 \hbar^4 m_e \omega} \Bigg[ 6 \hbar^2 M_C \eta_2 \cos\psi \notag \\ {} &- \mathcal{A}_0^2 e^2 \Big( M_C \big[ \mathcal{S}^2(12\eta_1 - 3\eta_2) - 16\eta_1 + 22\eta_2 \big] \cos\psi + 4 M_S (1-\eta_1\eta_2)(4 - 3\mathcal{S}^2) \sin\psi \Big) \Bigg],
		\end{align}
		and the $z$-component evaluates as
		\begin{equation}
			h_z^{(g)} = M_{\mathbf{k}}^{\rm g} - \frac{\mathcal{A}_0^2 e^2 \lambda_{\rm R}^2}{8\hbar^3 m_e \omega} \big( 2\eta_1 + \eta_2 \mathcal{S}^2 \big) + \frac{1}{8\hbar^3 m_e \omega} \Big( \Delta h_{z, 2}^{(g)} + \Delta h_{z, 3}^{(g)} + \Delta h_{z, 4}^{(g)} \Big).
		\end{equation}In this case, in-plane $\textbf{k}$-independent Zeeman-like fields, which break the $\mathcal{C}_2$ rotational symmetry, are also present.

		\section{S3. Effective Hamiltonian of 2D Rashba $\ell$-wave ($\ell=p,d,f,g$) magnets driven by two-color $\omega$--$2\omega$ hybrid linealry-circularly polarized light fields}
		In this section, we detail the effective Floquet Hamiltonians for a two-band system with varying AM symmetries driven by a hybrid circularly-linearly polarized light field. The external optical field is defined by the hybrid vector potentials
		\begin{subequations}
			\begin{align}
				A_x(t) &= \mathcal{A}_0 \big[ \cos(\omega t) + \mathcal{S}\cos(2\omega t + \psi)\cos\psi \big], \\
				A_y(t) &= \mathcal{A}_0 \big[ \eta_1 \sin(\omega t) + \mathcal{S}\cos(2\omega t + \psi)\sin\psi \big],
			\end{align}
		\end{subequations}
		where $\omega$ is the fundamental driving frequency, $\mathcal{S}$ controls the relative amplitude of the second harmonic, $\psi$ is the phase delay, $\psi$ is the polarization angle of the linear component, and $\eta_1 \in \{\pm 1\}$ determines the helicity of the fundamental circular field. Below are the fully expanded analytical components for $s$-, $p$-, $f$-, and $g$-wave AMs.
		
		\subsection*{S3.1. Isotropic Ponderomotive shift (common term)}
		
		Again, the identity-matrix coefficient $h_0(\mathbf{k})$ captures the unperturbed kinetic energy and the time-averaged ponderomotive shift, given by
		\begin{equation}
			h_0(\mathbf{k}) = \varepsilon(k_x^2 + k_y^2) + \frac{\varepsilon e^2 \mathcal{A}_0^2}{2\hbar^2}(2 +\mathcal{S}^2).
		\end{equation}
		
		\subsection*{S3.2. s-wave magnets}
		
		For an $s$-wave order parameter $M_{\mathbf{k}}^{\rm s} = M \hbar^2/m_e$, the light field induces a pure, momentum-independent effective out-of-plane Zeeman shift from the circularly polarized fundamental beam. The components are:
		\begin{subequations}
			\begin{align}
				h_x^{(s)} &= -\lambda_{\rm R} k_y, \\
				h_y^{(s)} &= \lambda_{\rm R} k_x, \\
				h_z^{(s)} &= M \frac{\hbar^2}{m_e} - \frac{\mathcal{A}_0^2 e^2 \eta_1 \lambda_{\rm R}^2}{\hbar^3 \omega}.
			\end{align}
		\end{subequations}Similar to the two other scenarios, there are no in-plane $\textbf{k}$-independent Zeeman-like fields $\mathcal{B}_x$ and $\mathcal{B}_y$ present to break the $\mathcal{C}_2$ rotational symmetry.
		
		\subsection*{S3.3. p-wave magnets}
		
		For a $p$-wave state $M_{\mathbf{k}}^{\rm p} = \frac{\hbar^2}{m_e}(M_C k_x + M_S k_y)$, the effective Floquet Hamiltonian components reveal light-induced momentum-independent pseudo-spin shifts indicative of a persistent bulk current as
		\begin{subequations}
			\begin{align}
				h_x^{(p)} &= -\lambda_{\rm R} k_y - \frac{\lambda_{\rm R} \mathcal{A}_0^2 e^2 \eta_1 M_S}{\hbar m_e \omega}, \\
				h_y^{(p)} &= \lambda_{\rm R} k_x + \frac{\lambda_{\rm R} \mathcal{A}_0^2 e^2 \eta_1 M_C}{\hbar m_e \omega}, \\
				h_z^{(p)} &= M_{\mathbf{k}}^{\rm p} - \frac{\mathcal{A}_0^2 e^2 \eta_1 \lambda_{\rm R}^2}{\hbar^3 m_e \omega}.
			\end{align}
		\end{subequations}Thus, there are in-plane $\textbf{k}$-independent Zeeman-like fields to break the $\mathcal{C}_2$ rotational symmetry.
		
		\subsection*{S4.4. f-wave magnets}
		
		For an $f$-wave AM state, the pristine exchange field is $M_{\mathbf{k}}^{\rm f} = \frac{\hbar^2}{m_e} \big[ M_C k_x(k_x^2 - 3k_y^2) + M_S k_y(k_y^2 - 3k_x^2) \big]$. Under the hybrid driving field, we can factor the dynamical corrections by powers of the field amplitude $\mathcal{A}_0$:
		\begin{subequations}
			\begin{align}
				h_x^{(f)} &= -\lambda_{\rm R} k_y + \frac{3 \lambda_{\rm R} \mathcal{A}_0^2 e^2 \eta_1}{\hbar m_e \omega} \big[ M_S(k_x^2 - k_y^2) + 2 M_C k_x k_y \big] \notag\\
				&\quad + \frac{3 \lambda_{\rm R} \mathcal{A}_0^4 \mathcal{S}^2 e^4 \eta_1}{2 \hbar^3 m_e \omega} \big[ M_S \cos(2\psi) + M_C \sin(2\psi) \big] \notag\\
				&\quad + \frac{3 \lambda_{\rm R} \mathcal{A}_0^3 \mathcal{S} e^3}{2 \hbar^2 m_e \omega} \Big[ \sin\psi \big( M_C k_x \cos\psi - M_S k_y \cos\psi - 2 M_C k_y \sin\psi - 2 M_S k_x \sin\psi \big) \notag\\
				&\hspace{2.8cm} - \eta_1 \cos\psi \big( M_C k_y \cos\psi + M_S k_x \cos\psi + 2 M_C k_x \sin\psi - 2 M_S k_y \sin\psi \big) \Big],\\
				h_y^{(f)} &= \lambda_{\rm R} k_x + \frac{3 \lambda_{\rm R} \mathcal{A}_0^2 e^2 \eta_1}{\hbar m_e \omega} \big[ M_C(k_x^2 - k_y^2) - 2 M_S k_x k_y \big] \notag\\
				&\quad + \frac{3 \lambda_{\rm R} \mathcal{A}_0^4 \mathcal{S}^2 e^4 \eta_1}{2 \hbar^3 m_e \omega} \big[ M_C \cos(2\psi) - M_S \sin(2\psi) \big] \notag\\
				&\quad + \frac{3 \lambda_{\rm R} \mathcal{A}_0^3 \mathcal{S} e^3}{2 \hbar^2 m_e \omega} \Big[ \sin\psi \big( 2 M_C k_y \cos\psi + 2 M_S k_x \cos\psi + M_C k_x \sin\psi - M_S k_y \sin\psi \big)\notag \\
				&\hspace{2.8cm} + \eta_1 \cos\psi \big( 2 M_C k_x \cos\psi - 2 M_S k_y \cos\psi - M_C k_y \sin\psi - M_S k_x \sin\psi \big) \Big],\\
				h_z^{(f)} &= M_{\mathbf{k}}^{\rm f} - \frac{\mathcal{A}_0^2 e^2 \eta_1 \lambda_{\rm R}^2}{\hbar^3 m_e \omega} \notag\\
				&\quad + \frac{3 \mathcal{A}_0^2 \mathcal{S}^2 e^2}{2 m_e} \Big[ \cos(2\psi) (M_C k_x - M_S k_y) - \sin(2\psi) (M_C k_y + M_S k_x) \Big] \notag\\
				&\quad + \frac{3 \mathcal{A}_0^3 \mathcal{S} e^3}{2 \hbar m_e} \Big[ \cos\psi (M_C \cos\psi - M_S \sin\psi) + \eta_1 \sin\psi (M_S \cos\psi + M_C \sin\psi) \Big].
			\end{align}
		\end{subequations}In this case, we also observe in-plane $\textbf{k}$-independent Zeeman-like fields.
		
		\subsection*{S3.5. g-wave magnets}
		For the $g$-wave order parameter $M_{\mathbf{k}}^{\rm g} = \frac{\hbar^2}{m_e} \big[ M_C(k_x^4 - 6k_x^2 k_y^2 + k_y^4) + 4M_S k_x k_y(k_x^2 - k_y^2) \big]$, the effective pseudo-spin components $h_x^{(g)}$ and $h_y^{(g)}$ expand as:
		\begin{subequations}
			\begin{align}
				h_x^{(g)} &= -\lambda_{\rm R} k_y - \frac{\lambda_{\rm R}}{8\hbar^4 m_e \omega} \Big( \Delta h_{x, 2}^{(g)} + \Delta h_{x, 3}^{(g)} + \Delta h_{x, 4}^{(g)} + \Delta h_{x, 5}^{(g)} \Big), \\
				h_y^{(g)} &= \lambda_{\rm R} k_x + \frac{\lambda_{\rm R}}{8\hbar^4 m_e \omega} \Big( \Delta h_{y, 2}^{(g)} + \Delta h_{y, 3}^{(g)} + \Delta h_{y, 4}^{(g)} + \Delta h_{y, 5}^{(g)} \Big),
			\end{align}
		\end{subequations}
		where the light-induced corrections, grouped by powers of the vector potential $\mathcal{A}_0$, are
		\begin{subequations}
			\begin{align}
				\Delta h_{x, 2}^{(g)} &= 16 \mathcal{A}_0^2 e^2 \eta_1 \hbar^3 \Big[ M_C k_y + 2 M_S k_x^3 - 6 M_C k_x^2 k_y - 6 M_S k_x k_y^2 \Big], \\
				\Delta h_{x, 3}^{(g)} &= 2 \mathcal{A}_0^3 \mathcal{S} e^3 \hbar^2 \Big[ \cos\psi \sin\psi \big( -M_C - 12 M_C k_x^2 + 18 M_C k_y^2 - 24 M_S k_x k_y \big) \notag \\
				&\quad + \sin\psi \sin\psi \big( -24 M_S k_x^2 + 24 M_S k_y^2 + 48 M_C k_x k_y \big) \notag \\
				&\quad + \eta_1 \cos\psi \cos\psi \big( 15 M_S - 18 M_S k_x^2 + 12 M_S k_y^2 + 24 M_C k_x k_y \big) \notag \\
				&\quad + \eta_1 \cos\psi \sin\psi \big( -4 M_C + 24 M_C k_x^2 + 48 M_S k_x k_y \big) \Big],  \\
				\Delta h_{x, 4}^{(g)} &= 24 \mathcal{A}_0^4 e^4 \eta_1 \hbar \Big[ -M_C k_y - 2 M_S \mathcal{S}^2 k_x - 2 M_C \mathcal{S}^2 k_y \cos^2\psi + 4 M_S \mathcal{S}^2 k_x \cos^2\psi \notag \\
				&\quad - 2 M_C \mathcal{S}^2 k_x \sin(2\psi) - 2 M_S \mathcal{S}^2 k_y \sin(2\psi) \Big], \\
				\Delta h_{x, 5}^{(g)} &= 3 \mathcal{A}_0^5 \mathcal{S} e^5 \Big[ \cos\psi \sin\psi \big( 6 M_C + 11 M_C \mathcal{S}^2 - 13 M_C \mathcal{S}^2 \cos^2\psi \big) \notag \\
				&\quad + \sin\psi \sin\psi \big( 16 M_S - 8 M_S \mathcal{S}^2 + 12 M_S \mathcal{S}^2 \sin^2\psi + 8 M_C \mathcal{S}^2 \cos^2\psi \cos\psi \big) \notag \\
				&\quad + \eta_1 \cos\psi \cos\psi \big( 4 M_S + 12 M_S \mathcal{S}^2 - 12 M_S \mathcal{S}^2 \cos^2\psi \big) \Big],\\
				\Delta h_{y, 2}^{(g)} &= 32 \mathcal{A}_0^2 e^2 \eta_1 \hbar^3 \Big[ M_C k_x^3 - M_S k_y^3 - 3 M_C k_x k_y^2 + 3 M_S k_x^2 k_y \Big], \\
				\Delta h_{y, 3}^{(g)} &= 6 \mathcal{A}_0^3 \mathcal{S} e^3 \hbar^2 \Big[ \cos\psi \sin\psi \big( -2 M_S - 2 M_S k_x^2 + 8 M_S k_y^2 + 16 M_C k_x k_y \big) \notag \\
				&\quad + \sin\psi \sin\psi \big( -M_C + 4 M_C k_x^2 + 2 M_C k_y^2 + 8 M_S k_x k_y \big) \notag \\
				&\quad + \eta_1 \cos\psi \cos\psi \big( -4 M_C + 8 M_C k_x^2 - 8 M_C k_y^2 + 16 M_S k_x k_y \big) \notag \\
				&\quad + \eta_1 \cos\psi \sin\psi \big( 2 M_S + 4 M_S k_x^2 - 4 M_S k_y^2 - 8 M_C k_x k_y \big) \Big], \\
				\Delta h_{y, 4}^{(g)} &= 24 \mathcal{A}_0^4 e^4 \eta_1 \hbar \Big[ -M_C k_x - M_S k_y - 2 M_C \mathcal{S}^2 k_x - 2 M_S \mathcal{S}^2 k_y + 4 M_C \mathcal{S}^2 k_x \cos^2\psi \notag \\
				&\quad + 4 M_S \mathcal{S}^2 k_y \cos^2\psi - 2 M_C \mathcal{S}^2 k_y \sin(2\psi) + 2 M_S \mathcal{S}^2 k_x \sin(2\psi) \Big], \\
				\Delta h_{y, 5}^{(g)} &= \mathcal{A}_0^5 \mathcal{S} e^5 \Big[ \cos\psi \sin\psi \big( -16 M_S + 24 M_S \mathcal{S}^2 - 36 M_S \mathcal{S}^2 \cos^2\psi \big) \notag \\
				&\quad + \sin\psi \sin\psi \big( 22 M_C + 30 M_C \mathcal{S}^2 - 27 M_C \mathcal{S}^2 \sin^2\psi \big) \notag \\
				&\quad + \eta_1 \cos\psi \cos\psi \big( 16 M_C - 24 M_C \mathcal{S}^2 + 36 M_C \mathcal{S}^2 \cos^2\psi \big) \notag \\
				&\quad + \eta_1 \cos\psi \sin\psi \big( 4 M_S - 6 M_S \mathcal{S}^2 + 36 M_S \mathcal{S}^2 \cos^2\psi \big) \Big].
			\end{align}
		\end{subequations}
		The $z$-component of the $g$-wave evaluates to
		\begin{align}
			h_z^{(g)} &= M_{\mathbf{k}}^{\rm g} - \frac{\mathcal{A}_0^2 e^2 \eta_1 \lambda_{\rm R}^2}{\hbar^3 m_e \omega} + \frac{\mathcal{A}_0^2 e^2}{8\hbar^3 m_e \omega} \Big[ 4 M_C \hbar^3 \omega (1 + \mathcal{S}^2) - 24 M_C \hbar^3 k_y^2 \omega - 48 M_S \mathcal{S}^2 \hbar^3 k_x k_y \omega \notag \\
			&\quad + 12 \mathcal{A}_0^2 M_C \mathcal{S}^2 e^2 \hbar \omega - 3 \mathcal{A}_0^2 M_C e^2 \hbar \omega + 21 \mathcal{A}_0^2 M_C \mathcal{S}^4 e^2 \hbar \omega \cos^4\psi - 6 \mathcal{A}_0^2 M_S \mathcal{S}^4 e^2 \hbar \omega \sin(2\psi) \notag \\
			&\quad - 4 M_C \mathcal{S}^2 \hbar^3 \omega \cos^2\psi + 24 \mathcal{A}_0^2 M_S \mathcal{S}^4 e^2 \hbar \omega \cos^3\psi \sin\psi \notag \\
			&\quad + 24 \mathcal{S}^2 \hbar^3 \omega \big( 2 M_C k_x^2 \cos^2\psi - M_C k_y^2 \cos^2\psi + M_S k_x^2 \sin(2\psi) - M_S k_y^2 \sin(2\psi) \notag \\
			&\quad - 2 M_C k_x k_y \sin(2\psi) + 4 M_S k_x k_y \cos^2\psi \big) \Big] \notag \\
			&\quad + \frac{3 \mathcal{A}_0^3 \mathcal{S} e^3}{8 \hbar m_e} \Big[ 16 M_C \hbar^2 k_x \cos\psi \cos\psi + 16 M_S \hbar^2 k_y \cos\psi \cos\psi - 8 M_C \hbar^2 k_y \cos\psi \sin\psi \notag \\
			&\quad + 16 M_S \hbar^2 k_x \cos\psi \sin\psi + 16 \eta_1 M_C \hbar^2 k_y \cos\psi \sin\psi - 16 \eta_1 M_S \hbar^2 k_x \cos\psi \sin\psi \notag \\
			&\quad + 16 \eta_1 M_C \hbar^2 k_x \sin\psi \sin\psi + 16 \eta_1 M_S \hbar^2 k_y \sin\psi \sin\psi \Big].
		\end{align}Interestingly, we also observe the footprint of in-plane $\textbf{k}$-independent Zeeman-like fields.

\begin{table*}[t]
			\centering
			\caption{\textbf{Rashba $d$-wave} altermagnet with $\omega$--$n\omega$ \textbf{linear--linear} driving with the N\'eel vector along the $z$-direction.}
			\renewcommand{\arraystretch}{1.5}
			\resizebox{1\linewidth}{!}{	\begin{tabular}{c|c|c|c|c|c|c}
					\hline\hline
					$n$ & Constant terms & $k_x^0k_y^1$ & $k_x^0k_y^2$ & $k_x^1k_y^0$ & $k_x^1k_y^1$ & $k_x^2k_y^0$ \\
					\hline
					
					1 &
					$\begin{aligned}
						h_0^{(0)} &= \frac{\mathcal{A}_0^2 e^2}{4m_e}
						\Big[
						1+\mathcal{S}^2 +2 \cos \beta \cos \psi 
						\Big]\\
						h_x^{(0)} &= 0
						\\
						h_y^{(0)} &= 0
						\\
						h_z^{(0)} &=
						\frac{\mathcal{A}_0^2 e^2}{4m_e}
						\left[
						M_C(1-\mathcal{S}^2+2\mathcal{S}^2\cos^2\beta)
						+2M_S\mathcal{S}^2\cos\beta\sin\beta
						\right]
						\\
						&\quad
						+\frac{\mathcal{A}_0^2 \mathcal{S} e^2\cos\psi}{2m_e}
						\left[
						M_C\cos\beta
						+M_S\sin\beta
						\right]
						\\
						&\quad
						+\frac{\mathcal{A}_0^2 \mathcal{S} e^2\lambda_R^2
							\sin\beta\sin\psi}
						{\hbar^3\omega}
					\end{aligned}$
					&
					$\begin{aligned}
						h_x &=
						-k_y\lambda_R
						\left[
						1+
						\frac{\mathcal{A}_0^2 M_C \mathcal{S} e^2
							\sin\beta\sin\psi}
						{\hbar m_e\omega}
						\right]
						\\[2mm]
						h_y &=
						-\frac{\mathcal{A}_0^2 M_S \mathcal{S} e^2
							k_y\lambda_R
							\sin\beta\sin\psi}
						{\hbar m_e\omega}
					\end{aligned}$
					&
					$\begin{aligned}
						h_z &=
						-\frac{M_C\hbar^2k_y^2}
						{2m_e}
					\end{aligned}$
					&
					$\begin{aligned}
						h_x &=
						\frac{\mathcal{A}_0^2 M_S \mathcal{S} e^2
							k_x\lambda_R
							\sin\beta\sin\psi}
						{\hbar m_e\omega}
						\\[2mm]
						h_y &=
						k_x\lambda_R
						\left[
						1-
						\frac{\mathcal{A}_0^2 M_C \mathcal{S} e^2
							\sin\beta\sin\psi}
						{\hbar m_e\omega}
						\right]
					\end{aligned}$
					&
					$\begin{aligned}
						h_z &=
						\frac{M_S\hbar^2k_xk_y}
						{m_e}
					\end{aligned}$
					&
					$\begin{aligned}
						h_z &=
						\frac{M_C\hbar^2k_x^2}
						{2m_e}
					\end{aligned}$
					\\
					\hline
					
					2 &
					$\begin{aligned}
						h_0^{(0)} &= \frac{\mathcal{A}_0^2 e^2}{4m_e}
						\Big[
						1+\mathcal{S}^2 
						\Big]\\
						h_x^{(0)} &=
						\frac{\mathcal{A}_0^3 \mathcal{S} e^3\lambda_R
							\sin\psi}
						{8\hbar^2m_e\omega}
						\left(
						3M_C\cos\beta
						+4M_S\sin\beta
						\right)
						\\[2mm]
						h_y^{(0)} &=
						-\frac{\mathcal{A}_0^3 M_C \mathcal{S} e^3\lambda_R
							\sin\beta\sin\psi}
						{8\hbar^2m_e\omega}
						\\[2mm]
						h_z^{(0)} &=
						\frac{\mathcal{A}_0^2 e^2}{4m_e}
						\left[
						M_C(1-\mathcal{S}^2+2\mathcal{S}^2\cos^2\beta)
						+2M_S\mathcal{S}^2\cos\beta\sin\beta
						\right]
					\end{aligned}$
					&
					$\begin{aligned}
						h_x &= -k_y\lambda_R
					\end{aligned}$
					&
					$\begin{aligned}
						h_z &= -\frac{M_C\hbar^2k_y^2}{2m_e}
					\end{aligned}$
					&
					$\begin{aligned}
						h_y &= k_x\lambda_R
					\end{aligned}$
					&
					$\begin{aligned}
						h_z &= \frac{M_S\hbar^2k_xk_y}{m_e}
					\end{aligned}$
					&
					$\begin{aligned}
						h_z &= \frac{M_C\hbar^2k_x^2}{2m_e}
					\end{aligned}$
					\\
					\hline
					
					$n\ge3$ &
					$\begin{aligned}
						h_0^{(0)} &= \frac{\mathcal{A}_0^2 e^2}{4m_e}
						\Big[
						1+\mathcal{S}^2 
						\Big]\\
						h_x^{(0)} &=0
						\\[2mm]
						h_y^{(0)} &=0
						\\[2mm]
						h_z^{(0)} &=
						\frac{\mathcal{A}_0^2 e^2}{4m_e}
						\left[
						M_C(1-\mathcal{S}^2+2\mathcal{S}^2\cos^2\beta)
						+2M_S\mathcal{S}^2\cos\beta\sin\beta
						\right]
					\end{aligned}$
					&
					$\begin{aligned}
						h_x &= -k_y\lambda_R
					\end{aligned}$
					&
					$\begin{aligned}
						h_z &= -\frac{M_C\hbar^2k_y^2}{2m_e}
					\end{aligned}$
					&
					$\begin{aligned}
						h_y &= k_x\lambda_R
					\end{aligned}$
					&
					$\begin{aligned}
						h_z &= \frac{M_S\hbar^2k_xk_y}{m_e}
					\end{aligned}$
					&
					$\begin{aligned}
						h_z &= \frac{M_C\hbar^2k_x^2}{2m_e}
					\end{aligned}$
					\\
					\hline\hline
					
			\end{tabular}}
			\label{tab:effective_hamiltonian_terms}
		\end{table*}
		
		\section{S4. Effective Hamiltonian of 2D Rashba $d$-wave altermagnets driven by two-color $\omega$--$n\omega$ polarized light fields}
		In this section, we present the effective Hamiltonian of a Rashba $d$-wave altermagnet under two-color laser driving, organized by powers of momentum $k$. The momentum-independent terms $h^{(0)}_{x/y}$ correspond to the effective in-plane Zeeman fields $\mathcal{B}_{x/y}$, which break $\mathcal{C}_{2z}$ symmetry and generate out-of-plane Edelstein responses. We classify these light-induced terms for bi-linear~(Tab.~\ref{tab:effective_hamiltonian_terms}), bi-circular~(Tab.~\ref{tab:floquet_hamiltonian_simplified}), and mixed circular--linear driving~(Tab.~\ref{tab:floquet_hamiltonian_simplified}), as well as for different harmonics $n$ of the second beam. It is evident that only $n=2$ yields finite $\mathcal{B}_{x/y}$ for all laser polarizations.

		\begin{table*}[t]
			\centering
			\caption{\textbf{Rashba $d$-wave} altermagnet with $\omega$--$n\omega$ \textbf{circular--circular} driving with the N\'eel vector along the $z$-direction.}
			\renewcommand{\arraystretch}{1.5}
			\resizebox{1\linewidth}{!}{
				\begin{tabular}{c|c|c|c|c|c|c}
					\hline\hline
					$n$ & Constant terms & $k_x^0k_y^1$ & $k_x^0k_y^2$ & $k_x^1k_y^0$ & $k_x^1k_y^1$ & $k_x^2k_y^0$ \\
					\hline
					
					1 &
					$\begin{aligned}
						h_0^{(0)} &= \frac{\mathcal{A}_0^2 e^2}{4m_e}
						\Big[
						2+2\mathcal{S}^2 +\mathcal{S} \cos \psi + \mathcal{S}  \eta_1 \eta_2 \cos \psi 
						\Big]\\
						h_x^{(0)} &= 0\\
						\quad h_y^{(0)} &= 0 \\
						h_z^{(0)} &=\frac{\mathcal{A}_0^2 \mathcal{S} e^2}{2m_e}
						\left[
						M_C(1-\eta_1\eta_2)\cos\psi
						+M_S(\eta_2-\eta_1)\sin\psi
						\right]
						\\&-\frac{\mathcal{A}_0^2 e^2\lambda_R^2}{\hbar^3\omega}
						\left[
						\eta_1+\mathcal{S}^2\eta_2
						+2S(\eta_1+\eta_2)\cos\psi
						\right]
					\end{aligned}$
					&
					$\begin{aligned}
						h_x &=
						\frac{k_y\lambda_R}{\hbar m_e\omega}
						\left[
						\mathcal{A}_0^2 M_C e^2
						\left(
						\eta_1 + \mathcal{S}^2\eta_2 + S(\eta_1+\eta_2)\cos\psi
						\right)
						-\hbar m_e\omega
						\right] \\
						h_y &=
						\frac{\mathcal{A}_0^2 M_\mathcal{S} e^2 k_y \lambda_R}{\hbar m_e \omega}
						\left[
						(\eta_1 + \mathcal{S}^2 \eta_2)
						+ S(\eta_1+\eta_2)\cos\psi
						\right]
					\end{aligned}$
					&
					$\begin{aligned}
						h_z &= -\frac{M_C\hbar^2 k_y^2}{2m_e}
					\end{aligned}$
					&
					$\begin{aligned}
						h_x &=
						-\frac{\mathcal{A}_0^2 M_\mathcal{S} e^2 k_x \lambda_R}{\hbar m_e \omega}
						\left[
						(\eta_1 + \mathcal{S}^2\eta_2)
						+ S(\eta_1+\eta_2)\cos\psi
						\right] \\
						h_y &=
						\frac{k_x \lambda_R}{\hbar m_e \omega}
						\left[
						\hbar m_e \omega
						+ \mathcal{A}_0^2 M_C e^2
						\left(
						\eta_1 + \mathcal{S}^2 \eta_2 + S(\eta_1+\eta_2)\cos\psi
						\right)
						\right]
					\end{aligned}$
					&
					$\begin{aligned}
						h_z &= \frac{M_S\hbar^2 k_x k_y}{m_e}
					\end{aligned}$
					&
					$\begin{aligned}
						h_z &= \frac{M_C\hbar^2 k_x^2}{2m_e}
					\end{aligned}$
					\\
					
					\hline
					
					2 &
					$\begin{aligned}
						h_0^{(0)} &= \frac{\mathcal{A}_0^2 e^2}{2m_e}
						\Big[
						1+\mathcal{S}^2 
						\Big]\\
						h_x^{(0)} &= \frac{\mathcal{A}_0^3 \mathcal{S} e^3 \lambda_R}{4\hbar^2 m_e \omega}
						\Big[
						M_C \sin\psi \,(1-\eta_1\eta_2)
						+ M_S \cos\psi \,(\eta_1 - 2\eta_2)
						\Big]\\
						h_y^{(0)} &= \frac{\mathcal{A}_0^3 \mathcal{S} e^3 \lambda_R}{4\hbar^2 m_e \omega}
						\Big[
						- M_S \sin\psi \,(1-\eta_1\eta_2)
						+ M_C \cos\psi \,(2\eta_1-\eta_2)
						\Big]\\
						h_z^{(0)} &= -\frac{\mathcal{A}_0^2 e^2 \lambda_R^2}{\hbar^3 \omega}
						\left(
						\eta_1 + \frac{\mathcal{S}^2 \eta_2}{2}
						\right)
					\end{aligned}$
					&
					$\begin{aligned}
						h_x &= \frac{k_y\,\lambda_R}{2\hbar m_e \omega}
						\Big[
						M_C \mathcal{A}_0^2 e^2(\eta_2 \mathcal{S}^2 + 2\eta_1)
						- 2\hbar m_e \omega
						\Big]\\
						h_y &= \frac{\mathcal{A}_0^2 M_\mathcal{S} e^2\,k_y\,\lambda_R}{2\hbar m_e \omega}
						\Big[
						\eta_2 \mathcal{S}^2 + 2\eta_1
						\Big]
					\end{aligned}$
					&
					$\begin{aligned}
						h_z&= -\frac{M_C\hbar^2 k_y^2}{2m_e}
					\end{aligned}$
					&
					$\begin{aligned}
						h_x &=
						-\frac{\mathcal{A}_0^2 M_\mathcal{S} e^2\, k_x \lambda_R}{2\hbar m_e \omega}
						(\eta_2 \mathcal{S}^2 + 2\eta_1)\\
						h_y &=
						\frac{k_x \lambda_R}{2\hbar m_e \omega}
						\Big[
						M_C \mathcal{A}_0^2 e^2(\eta_2 \mathcal{S}^2 + 2\eta_1)
						+ 2\hbar m_e \omega
						\Big]
					\end{aligned}$
					&
					$\begin{aligned}
						h_z & = \frac{M_S\hbar^2 k_x k_y}{m_e}
					\end{aligned}$
					&
					$\begin{aligned}
						h_z& = \frac{M_C\hbar^2 k_x^2}{2m_e}
					\end{aligned}$
					\\
					
					\hline
					
					$n\ge3$ &
					$\begin{aligned}
						h_0^{(0)} &= \frac{\mathcal{A}_0^2 e^2}{2m_e}
						\Big[
						1+\mathcal{S}^2 
						\Big]\\
						h_x^{(0)} &= 0\\
						\quad h_y^{(0)} &= 0 \\
						h_z^{(0)} &=
						-\frac{\mathcal{A}_0^2 e^2\lambda_R^2}{\hbar^3\omega}
						\Big(\eta_1 + \frac{\mathcal{S}^2\eta_2}{n}\Big)
					\end{aligned}$
					&
					$\begin{aligned}
						h_x &=
						\frac{k_y\,\lambda_R}{n\hbar m_e \omega}
						\Big[
						M_C \mathcal{A}_0^2 e^2(\eta_2 \mathcal{S}^2 + n\eta_1)
						- n\hbar m_e \omega
						\Big]\\
						h_y &=
						\frac{\mathcal{A}_0^2 M_\mathcal{S} e^2\,k_y\,\lambda_R}{n\hbar m_e \omega}
						(\eta_2 \mathcal{S}^2 + n\eta_1)
					\end{aligned}$
					&
					$\begin{aligned}
						h_z &= -\frac{M_C\hbar^2 k_y^2}{2m_e}
					\end{aligned}$
					&
					$\begin{aligned}
						h_x &=
						-\frac{\mathcal{A}_0^2 M_\mathcal{S} e^2\, k_x \lambda_R}{n\hbar m_e \omega}
						(\eta_2 \mathcal{S}^2 + n\eta_1)\\
						h_y &=
						\frac{k_x \lambda_R}{n\hbar m_e \omega}
						\Big[
						M_C \mathcal{A}_0^2 e^2(\eta_2 \mathcal{S}^2 + n\eta_1)
						+ n\hbar m_e \omega
						\Big]
					\end{aligned}$
					&
					$\begin{aligned}
						h_z & = \frac{M_S\hbar^2 k_x k_y}{m_e}
					\end{aligned}$
					&
					$\begin{aligned}
						h_z & =\frac{M_C\hbar^2 k_x^2}{2m_e}
					\end{aligned}$
					\\
					
					\hline\hline
			\end{tabular}}
			\label{tab:floquet_hamiltonian_simplified}
		\end{table*}
		
		\begin{table*}[t]
			\centering
			\caption{\textbf{Rashba $d$-wave} altermagnet with $\omega$--$n\omega$ \textbf{circular--linear} driving with the N\'eel vector along the $z$-direction.}
			\renewcommand{\arraystretch}{1.5}
			\resizebox{1\linewidth}{!}{
				\begin{tabular}{c|c|c|c|c|c|c}
					\hline\hline
					$n$ & Constant terms & $k_x^0k_y^1$ & $k_x^0k_y^2$ & $k_x^1k_y^0$ & $k_x^1k_y^1$ & $k_x^2k_y^0$ \\
					\hline
					
					1 &
					$\begin{aligned}
						h_0^{(0)} &= \frac{\mathcal{A}_0^2 e^2}{4m_e}
						\Big[
						2+\mathcal{S}^2 + 2 \mathcal{S} \cos\beta \cos \psi - 2 \eta_1 \mathcal{S} \sin\beta \sin \psi
						\Big]\\
						h_x^{(0)} &= 0\\
						\quad h_y^{(0)} &= 0 \\
						h_z^{(0)} &=\frac{\mathcal{A}_0^2 e^2}{2m_e}
						\Big[
						M_C \mathcal{S}^2 \cos^2\beta
						- \frac{1}{2}M_C \mathcal{S}^2
						+ M_C S \cos\beta \cos\psi
						+ M_C S \eta_1 \sin\beta \sin\psi
						\Big]
						\\ &+ \frac{\mathcal{A}_0^2 e^2}{2m_e}
						\Big[
						M_S \mathcal{S}^2 \cos\beta \sin\beta
						+ M_S S \cos\psi \sin\beta
						- M_S S \eta_1 \cos\beta \sin\psi
						\Big]
						\\&+ \frac{\mathcal{A}_0^2 e^2 \lambda_R^2}{\hbar^3 \omega}
						\Big[
						- \eta_1
						+ S \sin\beta \sin\psi
						- S \eta_1 \cos\beta \cos\psi
						\Big]
					\end{aligned}$
					&
					$\begin{aligned}
						h_x &=
						-\frac{k_y\,\lambda_R}{\hbar m_e \omega}
						\Big[
						\hbar m_e \omega
						- \mathcal{A}_0^2 M_C e^2 \eta_1
						+ \mathcal{A}_0^2 M_C \mathcal{S} e^2 \sin\beta \sin\psi
						- \mathcal{A}_0^2 M_C \mathcal{S} e^2 \eta_1 \cos\beta \cos\psi
						\Big]
						\\
						h_y &=\frac{\mathcal{A}_0^2 M_\mathcal{S} e^2\,k_y\,\lambda_R}{\hbar m_e \omega}
						\Big[
						\eta_1
						- S \sin\beta \sin\psi
						+ S \eta_1 \cos\beta \cos\psi
						\Big]
						
					\end{aligned}$
					&
					$\begin{aligned}
						h_z &= -\frac{M_C\hbar^2 k_y^2}{2m_e}
					\end{aligned}$
					&
					$\begin{aligned}
						h_x &=-\frac{\mathcal{A}_0^2 M_\mathcal{S} e^2\, k_x \lambda_R}{\hbar m_e \omega}
						\Big[
						\eta_1
						- S \sin\beta \sin\psi
						+ S \eta_1 \cos\beta \cos\psi
						\Big]
						\\
						h_y &=\frac{k_x \lambda_R}{\hbar m_e \omega}
						\Big[
						\hbar m_e \omega
						+ \mathcal{A}_0^2 M_C e^2 \eta_1
						- \mathcal{A}_0^2 M_C \mathcal{S} e^2 \sin\beta \sin\psi
						+ \mathcal{A}_0^2 M_C \mathcal{S} e^2 \eta_1 \cos\beta \cos\psi
						\Big]
						
					\end{aligned}$
					&
					$\begin{aligned}
						h_z &= \frac{M_S\hbar^2 k_x k_y}{m_e}
					\end{aligned}$
					&
					$\begin{aligned}
						h_z &= \frac{M_C\hbar^2 k_x^2}{2m_e}
					\end{aligned}$
					\\
					
					\hline
					
					2 &
					$\begin{aligned}
						h_0^{(0)} &= \frac{\mathcal{A}_0^2 e^2}{4m_e}
						\Big[
						2+\mathcal{S}^2 
						\Big]\\
						h_x^{(0)} &= \frac{\mathcal{A}_0^3 \mathcal{S} e^3 \lambda_R}{4\hbar^2 m_e \omega}
						\Big[
						M_C \cos\beta \sin\psi
						- 2M_C \eta_1 \sin\beta \cos\psi
						+ 2M_S \sin\beta \sin\psi
						+ 2M_S \eta_1 \cos\beta \cos\psi
						\Big]\\
						h_y^{(0)} &= \frac{\mathcal{A}_0^3 \mathcal{S} e^3 \lambda_R}{4\hbar^2 m_e \omega}
						\Big[
						M_C \sin\beta \sin\psi
						+ 2M_C \eta_1 \cos\beta \cos\psi
						- 2M_S \cos\beta \sin\psi
						+ M_S \eta_1 \sin\beta \cos\psi
						\Big]\\
						h_z^{(0)} &= + \frac{\mathcal{A}_0^2 e^2}{4m_e}
						\Big[
						M_C \mathcal{S}^2 (2\cos^2\beta - 1)
						+ 2M_S \mathcal{S}^2 \sin\beta \cos\beta
						\Big]
						- \frac{\mathcal{A}_0^2 e^2 \eta_1 \lambda_R^2}{\hbar^3 \omega}
					\end{aligned}$
					&
					$\begin{aligned}
						h_x &=-k_y\,\lambda_R
						\left[
						1 - \frac{\mathcal{A}_0^2 M_C e^2 \eta_1}{\hbar m_e \omega}
						\right]\\
						h_y &=\frac{\mathcal{A}_0^2 M_\mathcal{S} e^2 \eta_1\, k_y\, \lambda_R}{\hbar m_e \omega}
					\end{aligned}$
					&
					$\begin{aligned}
						h_z&= -\frac{M_C\hbar^2 k_y^2}{2m_e}
					\end{aligned}$
					&
					$\begin{aligned}
						h_x &=-\frac{\mathcal{A}_0^2 M_\mathcal{S} e^2 \eta_1\, k_x \lambda_R}{\hbar m_e \omega}
						\\
						h_y &=k_x \lambda_R
						\left[
						1 + \frac{\mathcal{A}_0^2 M_C e^2 \eta_1}{\hbar m_e \omega}
						\right]
						
					\end{aligned}$
					&
					$\begin{aligned}
						h_z & = \frac{M_S\hbar^2 k_x k_y}{m_e}
					\end{aligned}$
					&
					$\begin{aligned}
						h_z& = \frac{M_C\hbar^2 k_x^2}{2m_e}
					\end{aligned}$
					\\
					
					\hline
					
					$n\ge3$ &
					$\begin{aligned}
						h_0^{(0)} &= \frac{\mathcal{A}_0^2 e^2}{4m_e}
						\Big[
						2+\mathcal{S}^2 
						\Big]\\
						h_x^{(0)} &= 0\\
						\quad h_y^{(0)} &= 0 \\
						h_z^{(0)} &=\frac{\mathcal{A}_0^2 M_C \mathcal{S}^2 e^2}{4m_e}
						\left(2\cos^2\beta - 1\right)
						+
						\frac{\mathcal{A}_0^2 M_S \mathcal{S}^2 e^2 \sin\beta \cos\beta}{2m_e}
						-
						\frac{\mathcal{A}_0^2 e^2 \eta_1 \lambda_R^2}{\hbar^3 \omega}
						
					\end{aligned}$
					&
					$\begin{aligned}
						h_x &=-k_y\,\lambda_R
						\left[
						1 - \frac{\mathcal{A}_0^2 M_C e^2 \eta_1}{\hbar m_e \omega}
						\right]
						\\
						h_y &=\frac{\mathcal{A}_0^2 M_\mathcal{S} e^2 \eta_1\, k_y\, \lambda_R}{\hbar m_e \omega}
						
					\end{aligned}$
					&
					$\begin{aligned}
						h_z &= -\frac{M_C\hbar^2 k_y^2}{2m_e}
					\end{aligned}$
					&
					$\begin{aligned}
						h_x &=-\frac{\mathcal{A}_0^2 M_\mathcal{S} e^2 \eta_1\, k_x \lambda_R}{\hbar m_e \omega}
						\\
						h_y &=k_x \lambda_R
						\left[
						1 + \frac{\mathcal{A}_0^2 M_C e^2 \eta_1}{\hbar m_e \omega}
						\right]
						
					\end{aligned}$
					&
					$\begin{aligned}
						h_z & = \frac{M_S\hbar^2 k_x k_y}{m_e}
					\end{aligned}$
					&
					$\begin{aligned}
						h_z & =\frac{M_C\hbar^2 k_x^2}{2m_e}
					\end{aligned}$
					\\
					
					\hline\hline
			\end{tabular}}
			\label{tab:floquet_hamiltonian_simplified}
		\end{table*}
		
	}
\end{document}